\documentclass[onecolumn,preprintnumbers,amsmath,aps]{revtex4}
\usepackage{graphicx}
\usepackage{dcolumn}
\usepackage{bm}
\usepackage{color}
\usepackage{multirow}
\begin{document}
\def\eq#1{(\ref{#1})}
\def\fig#1{\ref{#1}}
\def\tab#1{\ref{#1}}
\title{Interaction of the hydrogen molecule with the environment:\\ stability of the system}
\author{I. A. Wrona$^{\left(1\right)}$}
\author{M. W. Jarosik$^{\left(2\right)}$}
\author{R. Szcz{\c{e}}{\'s}niak$^{\left(2\right)}$}
\author{K. A. Szewczyk$^{\left(1\right)}$}
\author{M. K. Stala$^{\left(2\right)}$}
\author{W. Leo{\'n}ski$^{\left(3\right)}$}
\affiliation{$^1$ Institute of Physics, Jan D{\l}ugosz University in Cz{\c{e}}stochowa, Ave. Armii Krajowej 13/15, 42-200 Cz{\c{e}}stochowa, Poland}
\affiliation{$^2$ Institute of Physics, Cz{\c{e}}stochowa University of Technology, Ave. Armii Krajowej 19, 42-200 Cz{\c{e}}stochowa, Poland}
\affiliation{$^3$ Quantum Optics and Engineering Division, Faculty of Physics and Astronomy, University of Zielona G{\'o}ra, 
Prof. Z. Szafrana 4a, 65-516 Zielona G{\'o}ra, Poland}
\date{\today}
\begin{abstract}
We study the stability of the hydrogen molecule interacting with the environment according to the balanced gain 
and loss energy scheme. We determined the properties of the molecule taking into account all electronic interactions, 
where the parameters of the Hamiltonian have been computed by using the variational method. 
The interaction of the hydrogen molecule with the environment was modeled parametrically ($\gamma$) 
with the help of the non-hermitian operator. We have shown that the hydrogen molecule is dynamically unstable. 
The dissociation time ($T_{D}$) decreases, if the $\gamma$ parameter increases (for $\gamma\rightarrow 0$, 
we get $T_{D}\rightarrow +\infty$). 
At the dynamic instability of the hydrogen molecule overlaps its static instability as the coupling constant $\gamma$ increases. 
We observed the decrease in the dissociation energy and the existence of the metastable state of the molecule ($\gamma_{MS}=0.659374$~Ry). 
The hydrogen molecule is statically unstable for $\gamma >\gamma_{D}=1.024638$~Ry.
One can also observed the $\mathcal{PT}$ symmetry breaking effect for the electronic Hamiltonian ($\gamma_{\mathcal {PT}}=0.520873$~Ry). 
However, it does not affect the properties of the hydrogen molecule, such as: the electronic Hamiltonian parameters, the phonon and rotational energy, 
and the values of the electron-phonon coupling constants. 
\end{abstract}
\maketitle
{\bf Keywords:} Hydrogen molecule, Interaction with the environment, Dynamic and static stability,\\ $\mathcal{PT}$ symmetry breaking.
%

\section{Introduction}

Research on the impact of the environment (external quantum system) on the state of the quantum system is the interesting but very difficult issue \cite{Davies1976A, Breuer2007A}. This is due to two reasons: 
(i) usually in the case of the realistic quantum system it is impossible to accurately determine its internal state due to the complexity, and 
(ii) the interaction between the quantum system and the environment can be so complicated that it is impossible to obtain unambiguous results. 

In the paper, we took into account the hydrogen molecule, which interacts with the environment according to the {\bf B}alanced {\bf G}ain and {\bf L}oss (BGL) energy scheme \cite{Klett2017}. The hydrogen molecule is an interesting case in that it represents the non-trivial quantum system, and its state can be described accurately using the variational method \cite{Kolos1964A, Kolos1968A, Spalek2014A, Jarosik2018A}. On the other hand, the BGL scheme describes the interaction between the molecule and the environment in the realistic and simple way. From the mathematical point of view, the BGL type interaction is modeled by the non-hermitian Hamiltonian \cite{Bender2007, Moiseyev2011}. However, it is invariant due to the $\mathcal{PT}$ symmetry  (the symmetry of reflection in space ($\mathcal{P}$) and in time ($\mathcal{T}$)) \cite{Bender1998, Bender1999, Bender2002, Bender2003, Bender2008}. We underline that if the Hamiltonian is not-hermitian, but it has the unbroken $\mathcal{PT}$ symmetry, then the energy spectrum of the system is real - at least to the characteristic value of the parameter controlling the interaction with the environment. 

The interest in the non-hermitian Hamiltonians, in the context of the description of the open systems, appeared in many areas of physics.
It is worth mentioning the papers \cite{Hiller2006, Graefe2008A}, in which the open Bose-Hubbard dimmer was analyzed. 
Such system can be implemented experimentally in the form of trapped bosons, where the coupling constant between the studied system and 
the environment reflects the value of barrier potential \cite{Graefe2010A}. In the context of Bose-Einstein condensate the $\mathcal{PT}$ symmetry 
breaking was also analyzed in the works \cite{Graefe2012A, Kreibich2016A, Dast2017A}. It is worth noting that the existence of the $\mathcal{PT}$ symmetry breaking were confirmed in the field of quantum optics \cite{Makris2008A, Guo2009A, Ruter2010A, Peng2014A}. Additionaly, the complex energy values are used 
to explain the probability of disintegration of the physical systems, the transport mechanism or the dispersion 
\cite{Dattoli1990A, Moiseyev1998A, Okolowicz2003A, Berry2004A, Graefe2010B}. Although most of the existing models have been introduced heuristically,
 however on the basis of relatively satisfactory mathematical justification \cite{Graefe2010A}. 

On the basis of the discussed issues, we intend to analyze the hydrogen molecule interacting with the environment and to examine 
her stability. We assume that the required calculations will be carried out in the extremely accurate manner (at the level required in the quantum chemistry), so that the obtained results could be verified experimentally. In our opinion, the results presented in the paper may be helpful in the development of the quantum computer based on the hydrogen molecules, with each molecule considered as the carrier of qubits \cite{Setia2018A}. 
In particular, they can help to characterize the stability of the system.

\section{Formalism}

The total energy ($E_{T}$) of hydrogen molecule is defined as:
\begin{equation}
\label{r01-II}
E_{T}=E_{p}+E_{e\gamma},
\end{equation}
where: $E_{p}=2/R$ represents the energy of proton repulsion, with $R=|{\bf R}|$ as the distance between protons, 
       $E_{e\gamma}$ means the energy of the lowest electronic state in the presence of the loss and gain effect 
       ($\gamma$ represents the coupling between the molecule and the environment). 
       For $\gamma=0$, the energy $E_{e}=E_{e\left(\gamma=0\right)}$ should be determined using the Hubbard Hamiltonian, 
       which takes into account all electronic interactions. 
       In the second quantization formalism, we have \cite{Spalek2014A}:
\begin{eqnarray}
\label{r02-II}
\hat{\mathcal{H}_{e}}&=&\varepsilon\left(\hat{n}_1+\hat{n}_2\right)
       +t\sum_{\sigma}\left(\hat{c}_{1\sigma}^{\dag} \hat{c}_{2\sigma}+\hat{c}_{2\sigma}^{\dag}\hat{c}_{1\sigma}\right)
       +U\left(\hat{n}_{1\uparrow}\hat{n}_{1\downarrow}+\hat{n}_{2\uparrow}\hat{n}_{2\downarrow}\right)
       +\left(K-\frac{J}{2} \right)\hat{n}_1\hat{n}_2\\ \nonumber 
       &-&2J\hat{\mathbf{S}}_1\hat{\mathbf{S}}_2 
        +J\left(\hat{c}_{1\uparrow}^{\dag}\hat{c}_{1\downarrow}^{\dag} \hat{c}_{2\downarrow}\hat{c}_{2\uparrow}+h.c.\right) 
        +V\sum_{\sigma}\left[\left(\hat{n}_{1-\sigma}+\hat{n}_{2-\sigma} \right) \left(\hat{c}_{1\sigma}^{\dag} \hat{c}_{2\sigma}
        +\hat{c}_{2\sigma}^{\dag}\hat{c}_{1\sigma}\right)\right],
\end{eqnarray}
where the symbol $\hat{n}_{j}$ is given by: $\hat{n}_{j}=\sum_{\sigma}\hat{n}_{j\sigma}=\sum_{\sigma}\hat{c}^{\dag}_{j\sigma}\hat{c}_{j\sigma}$, and 
$\hat{c}^{\dag}_{j\sigma}$ ($\hat{c}_{j\sigma}$) is the electron creation (annihilation) operator, which refers to the $j$-th hydrogen atom, 
$\sigma$ represents the electronic spin: $\sigma\in\{\uparrow,\downarrow\}$. 
The product of spin operators $\hat{{\bf S}}_{i}\hat{{\bf S}}_{j}$ is in the form of: 
$\frac{1}{2}\left(\hat{S}^{+}_{i}\hat{S}^{-}_{j}+\hat{S}^{-}_{i}\hat{S}^{+}_{j}\right)+\hat{S}^{z}_{i}\hat{S}^{z}_{j}$, where $\hat{S}^{+}_{j}=\hat{c}_{j\uparrow}^{\dag}\hat{c}_{j\downarrow}$, $\hat{S}^{-}_{j}=\hat{c}_{j\downarrow}^{\dag}\hat{c}_{j\uparrow}$, and 
$\hat{S}^{z}_{j}=\frac{1}{2}\left(\hat{n}_{j\uparrow}-\hat{n}_{j\downarrow}\right)$. The Hamiltonian parameters are defined by the following integrals:
\begin{eqnarray}
\label{r03-II}
\varepsilon &=&
\int d^{3}{\bf r}\Phi_{1}\left({\bf r}\right)\left[-\nabla^{2}-\frac{2}{|{\bf r}-{\bf R}|}\right]\Phi_{1}\left({\bf r}\right),\\ \nonumber
t &=&
\int d^{3}{\bf r}\Phi_{1}\left({\bf r}\right)\left[-\nabla^{2}-\frac{2}{|{\bf r}-{\bf R}|}\right]\Phi_{2}\left({\bf r}\right),\\ \nonumber
U &=&
\int\int d^{3}{\bf r}_{1}d^{3}{\bf r}_{2}\Phi^{2}_{1}\left({\bf r}_{1}\right)\frac{2}{|{\bf r}_{1}-{\bf r}_{2}|}\Phi^{2}_{1}\left({\bf r}_{2}\right),\\ \nonumber
K &=&
\int\int d^{3}{\bf r}_{1}d^{3}{\bf r}_{2}\Phi^{2}_{1}\left({\bf r}_{1}\right)\frac{2}{|{\bf r}_{1}-{\bf r}_{2}|}\Phi^{2}_{2}\left({\bf r}_{2}\right),\\ \nonumber
J &=&
\int\int d^{3}{\bf r}_{1}d^{3}{\bf r}_{2}\Phi_{1}\left({\bf r}_{1}\right)\Phi_{2}\left({\bf r}_{1}\right)\frac{2}{|{\bf r}_{1}-{\bf r}_{2}|}\Phi_{1}\left({\bf r}_{2}\right)\Phi_{2}\left({\bf r}_{2}\right),\\ \nonumber
V &=&
\int\int d^{3}{\bf r}_{1}d^{3}{\bf r}_{2}\Phi^{2}_{1}\left({\bf r}_{1}\right)\frac{2}{|{\bf r}_{1}-{\bf r}_{2}|}\Phi_{1}\left({\bf r}_{1}\right)\Phi_{2}\left({\bf r}_{2}\right). 
\end{eqnarray}
The meaning of above quantities is as follows: $\varepsilon$ represents the energy of the molecular orbital, 
$t$ is the electronic hopping integral, $U$ denotes the on-site Coulomb repulsion, $K$ is the energy of the intersitial Coulomb repulsion, 
$J$ represents the integral of the exchange, and $V$ is called the correlated hopping. 
The integrals are calculated numerically, which is the complicated procedure that requires the use of the large computer resources. 
Let me note that the contribution of the individual integrals to the energy eigenvalues is very diversed (see table \tab{tA1} in the appendix \ref{dodA}), nevertheless omitting any interaction would lead to the non-physical shortening of the distance between protons. We chose the Wannier's functions in the form of:  
\begin{equation}
\label{r04-II}
\Phi_{j}\left({\bf r}\right)= a\left[\phi_{j}\left({\bf r}\right)-b\phi_{l}\left({\bf r}\right)\right],
\end{equation}
where the coefficients ensuring normalization are expressed in the formulas:
\begin{eqnarray}
\label{r05-II}
a=\frac{1}{\sqrt{2}}\sqrt{\frac{1+\sqrt{1-S^{2}}}{1-S^{2}}}, \hspace{5mm} b=\frac{S}{1+\sqrt{1-S^{2}}}.
\end{eqnarray}
The atomic overlap (S) should be calculated using the formula: 
$S=\int d^{3}{\bf r}\phi_{1}\left({\bf r}\right)\phi_{2}\left({\bf r}\right)$, 
where $1s$ Slater-type orbital can be written as: 
$\phi_{j}\left({\bf r}\right)=\sqrt{\alpha^{3}/\pi}\exp\left[-\alpha|{\bf r}-{\bf R}_{j}|\right]$, $\alpha$ is the inverse size of the orbital. 
It should be noted that the second quantization method is completely equivalent to the Schr{\"o}dinger analysis \cite{Schrodinger1926A, Schrodinger1926B, Schrodinger1926C, Schrodinger1926D, Fetter1971A}. 

The effective interaction of hydrogen molecule with the environment will be taken into account by supplementing the Hubbard Hamiltonian 
$\hat{\mathcal{H}_{e}}$ with the balanced gain and loss operator \cite{Klett2017, Dast2016}:
\begin{equation}
\label{r06-II}
\hat{\mathcal{H}}_{\gamma}=i\gamma\left(\hat{n}_{1}-\hat{n}_{2}\right).
\end{equation}

Adding the operator $\hat{\mathcal{H}}_{\gamma}$ to $\hat{\mathcal{H}}_{e}$ results in the loss of hermitism of full electronic Hamiltonian 
($\hat{\mathcal{H}}_{e\gamma}=\hat{\mathcal{H}}_{e}+\hat{\mathcal{H}}_{\gamma}$). However, it remains invariant due 
to the $\mathcal{PT}$ symmetry - at least to the characteristic $\gamma_{\mathcal{PT}}$ value. 
The easiest way to understand the physical significance of $\hat{\mathcal{H}}_{\gamma}$ is to refer to the case 
of the subsystem described by Hamiltonian $\hat{\mathcal{H}}$ to which added the constant imaginary contribution $i\Gamma$ \cite{Cartarius2014}. 
In the case at hand, the Schr{\"o}dinger equation has the form: 
$i\partial\Psi\left({\bf r},t\right)/\partial t =\left(\hat{\mathcal{H}}+i\Gamma\right)\Psi\left({\bf r},t\right)$.
The solution can be presented in the following form:
$\Psi\left({\bf r}, t \right)=\psi\left({\bf r} \right)\exp\left(-iEt \right)\exp\left(\Gamma t \right)$, 
where $\hat{\mathcal {H}}\psi\left({\bf r}\right)=E\psi\left({\bf r}\right)$. 
Taking into account 
$\rho\left({\bf r}, t\right)=|\Psi\left({\bf r}, t\right)|^{2}=|\psi\left({\bf r}\right)|^2\exp\left(2\Gamma t\right)$, 
we state that depending on the signum of $\Gamma$ the probability amplitude increases or decreases exponentially. The interpretation is that the particles described by the wave function enter or leave subsystem, wherein the total number of particles (subsystem and environment) being constant. 
Note that the full system is always described by the hermitian Hamiltonian.

\section{Results}
%

\subsection{Static stability of the system: the electron, phonon and electron-phonon properties}
\begin{figure*} 
\includegraphics[width=1\columnwidth]{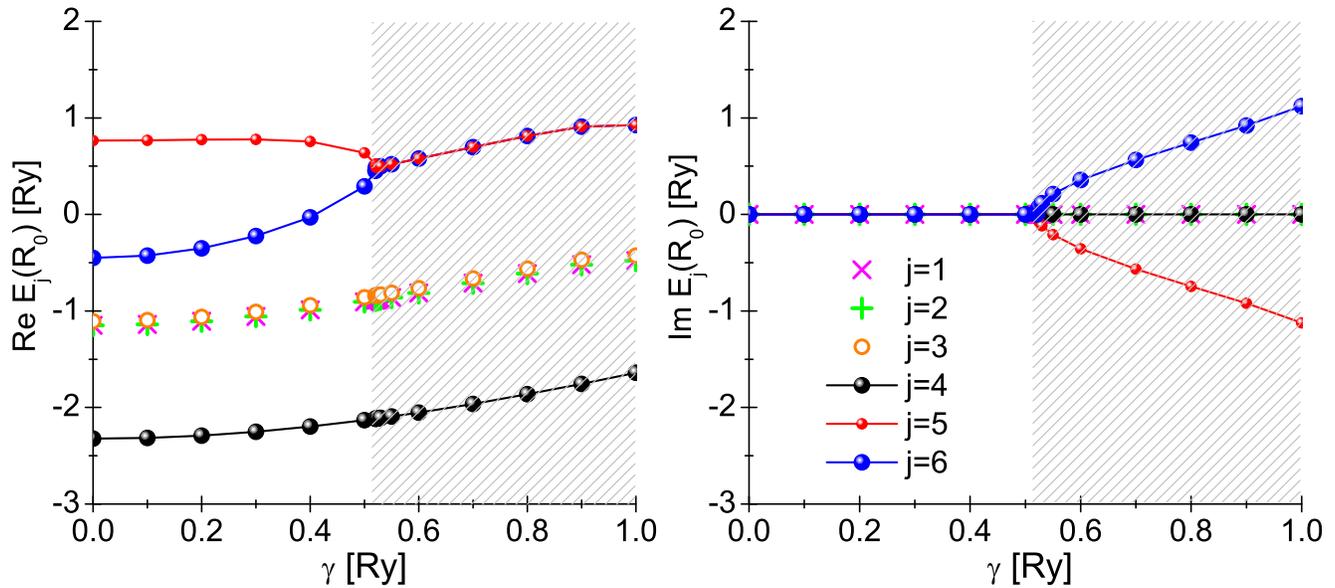}
\caption{The real and imaginary part of the eigenvalues of the Hamiltonian $\hat{\mathcal{H}}_{e\gamma}$. 
         We assume the equilibrium distances between protons ($R_{0}$). The hatched areas correspond to the $\gamma$ values 
         for which the operator $\hat{\mathcal{H}}_{e\gamma}$ ceases to be invariant due to the $\mathcal{PT}$ symmetry.}
\label{f1}
\end{figure*}

In the figure \fig{f1}, we plotted the dependency of the eigenvalues $E_ {j}$ on $\gamma$. 
The analytical formulas for $E_{j} $ have been collected in the Appendix~\ref{dodA}. Analyzing the obtained results, we found that for 
$\gamma_{\mathcal{PT}}=0.520873$~Ry there is the breaking of $\mathcal {PT}$ symmetry of the electronic Hamiltonian. 
This fact is manifested by the appearance of the complex values of $E_{5}$ and $E_{6}$. 
Physically, this means that the $\mathcal{PT}$ symmetry breaking reduces the number of the available electronic states from six to four. 
Nevertheless, the considered effect has no physical significance due to the fact that the states $|\left.E_{5}\right>$ and $|\left.E_{6}\right>$ have the highest energy values. They can not be thermally occuped - the $k_{B}T$ energy is of the order of $25$~meV, while the difference between $E_{6}$ and $E_{4}$ 
is around $25.5$~eV ($E_{4}$ is the ground state energy of the electronic subsystem). When discussing the results, it should be clearly emphasized that $E_{4}$ always accepts the real values.

\begin{figure*} 
\includegraphics[width=\columnwidth]{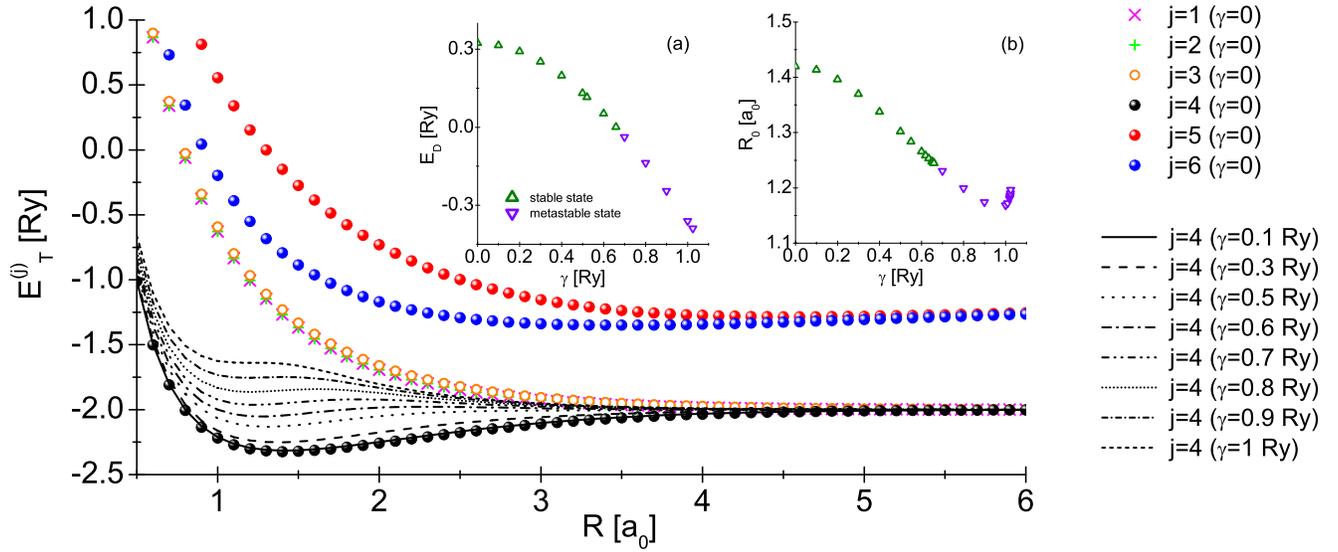}
\caption{    The dependence of the total energy $E_{T}^{\left(j\right)}$ on the distance between protons. 
             Additionaly, we take into account the influence of $\gamma$ on the ground state energy $E_{T}^{\left(4\right)}$.
             At the insert (a) - the dissociation energy $E_{D} $ as the function of $\gamma$.
             At the insert (b), we plot the dependence of the equilibrium distance $R_{0}$ on the $\gamma$ parameter.}  
                       
\label{f2}
\end{figure*}
\begin{figure*}
\includegraphics[width=0.3\columnwidth]{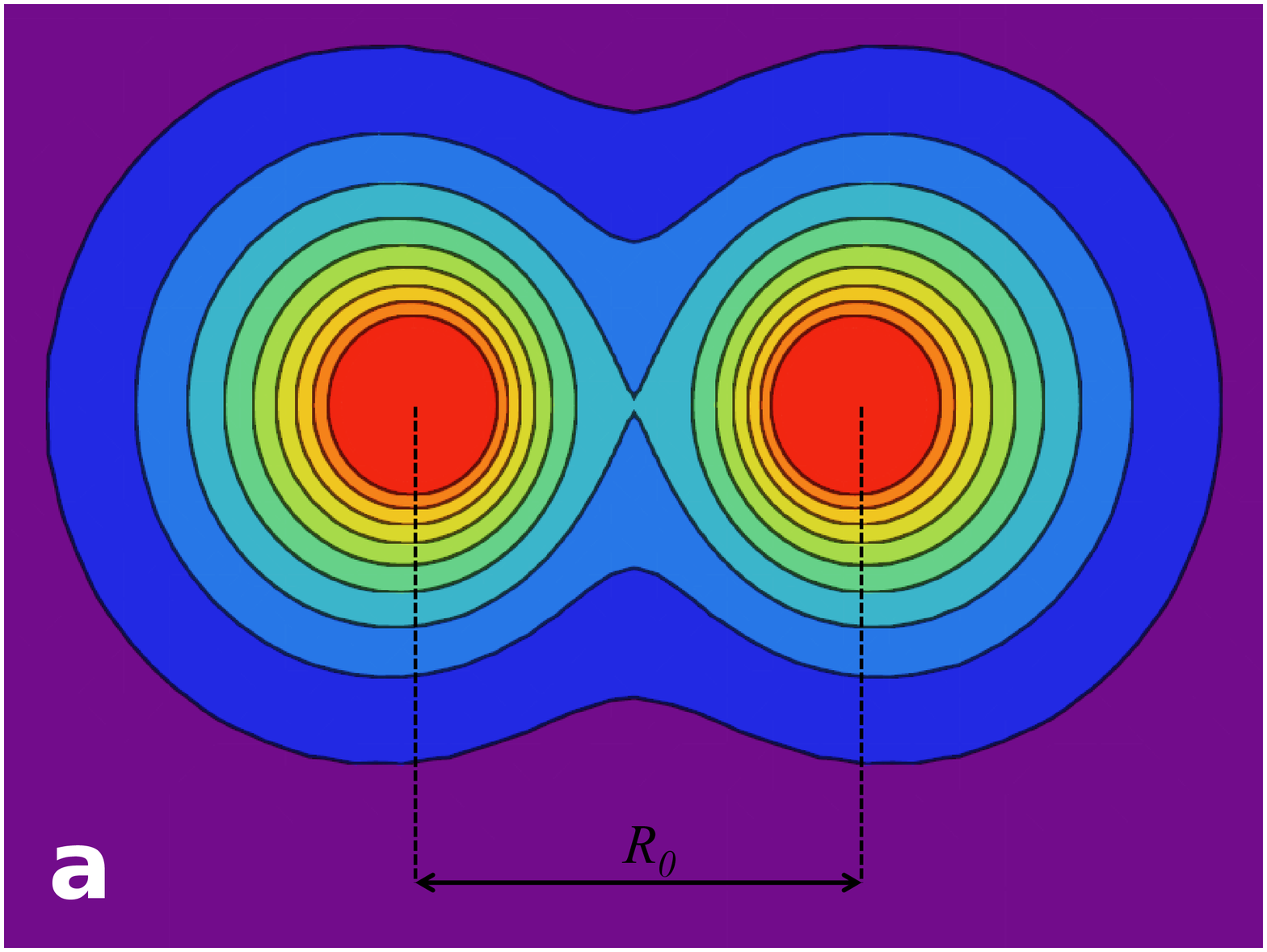}
\includegraphics[width=0.3\columnwidth]{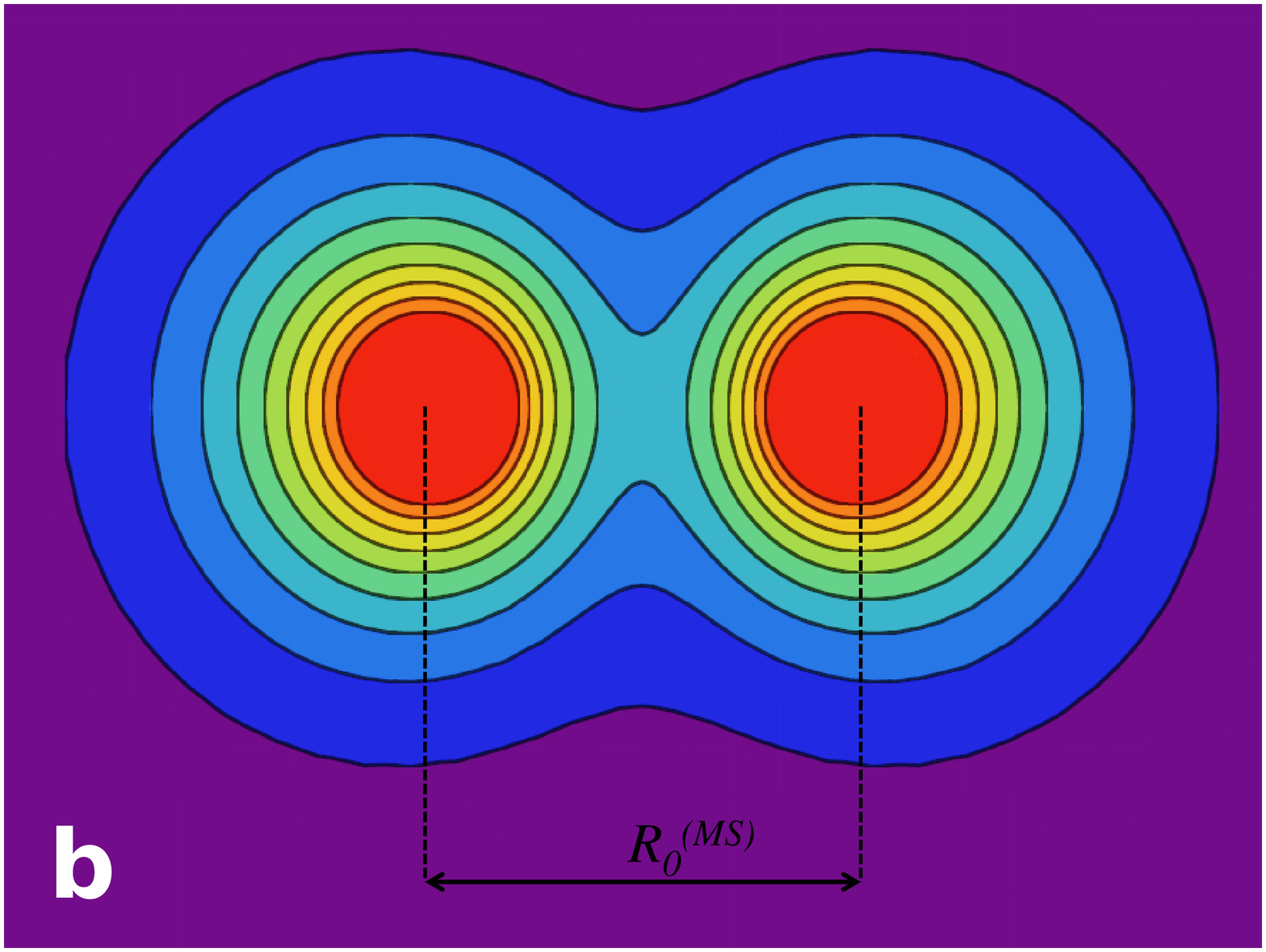}
\includegraphics[width=0.3\columnwidth]{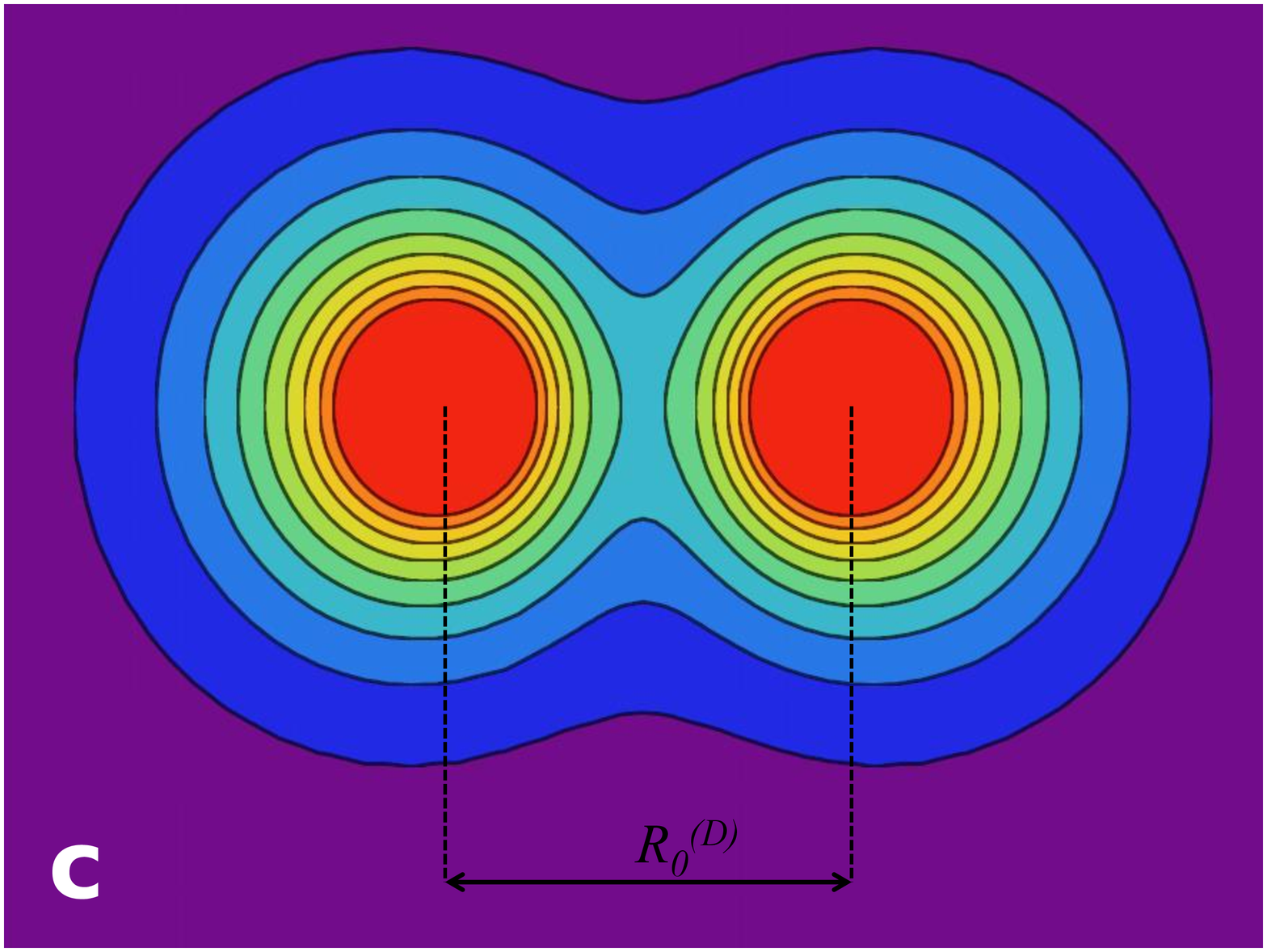}
\caption{The distribution of electronic charge in the hydrogen molecule 
         for the stable (a) and metastable (b) case, and just before disotiation (c).}           
\label{f3}
\end{figure*}
\begin{figure*}
\includegraphics[width=0.8\columnwidth]{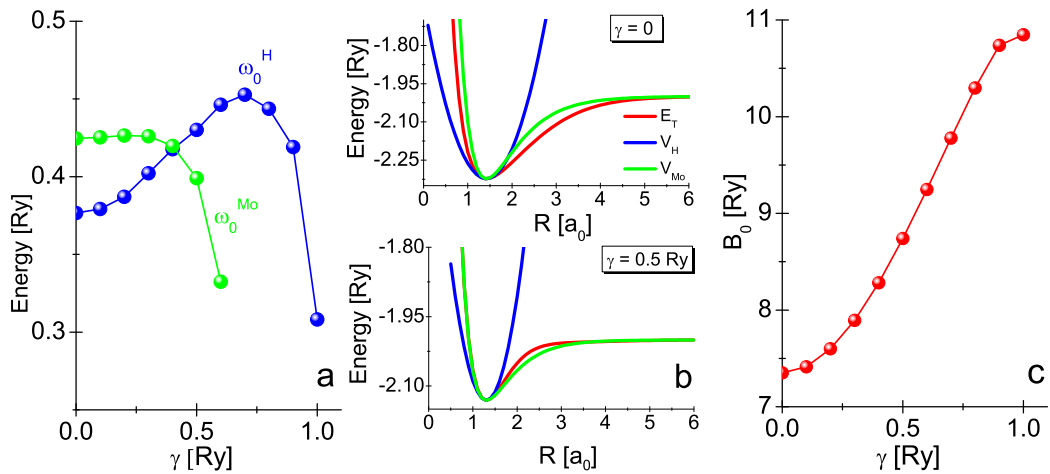}
\caption{(a) The influence of the $\gamma$ parameter on the energy values $\omega^{\rm H}_{0}$ and $\omega^{\rm Mo}_{0}$.
         (b) The exemplary parameterization of the total energy curve in the harmonic and anharmonic Morse case. 
         (c) The rotational constant $B_{0}$ as the $\gamma$ parameter function.}             
\label{f4}
\end{figure*}
\begin{figure*}
\includegraphics[width=\columnwidth]{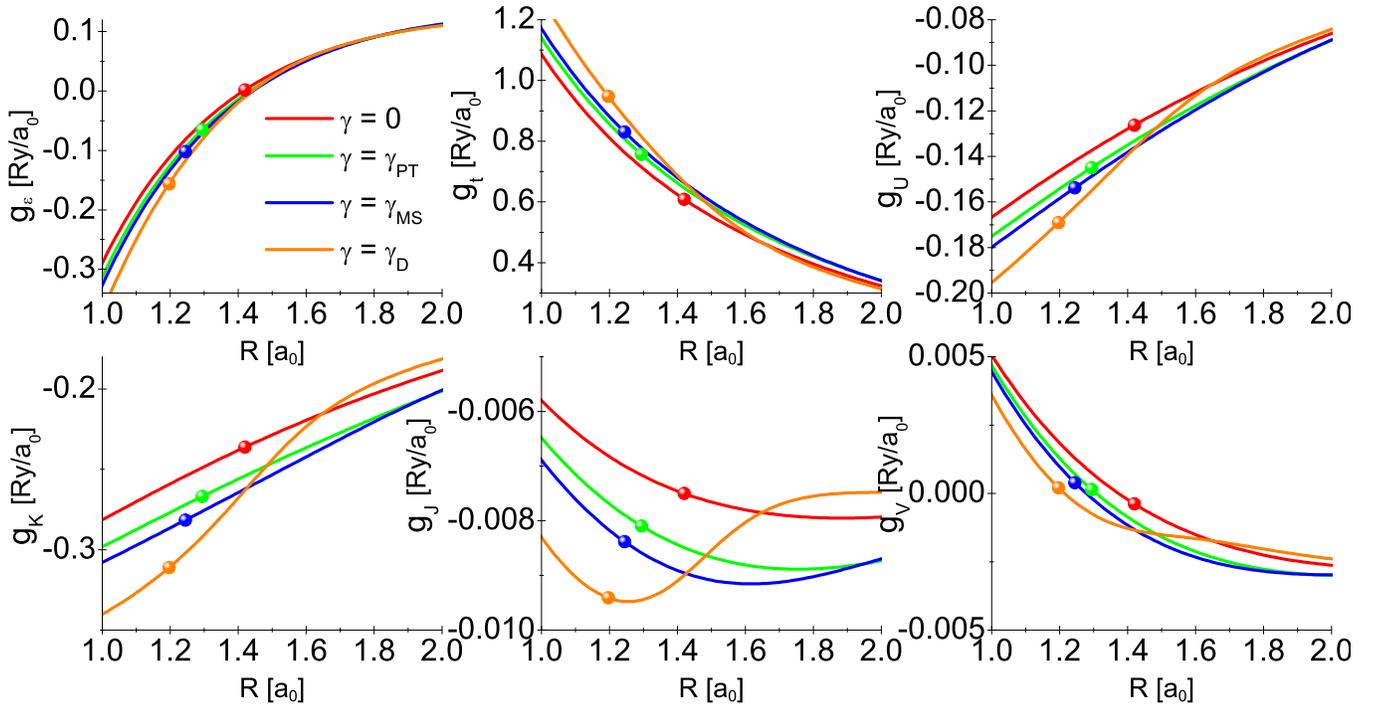}
\caption{The electron-phonon coupling as a function of inter-proton distance for selected values of the $\gamma$ parameter. 
         The symbols placed on the curves point to the equilibrium value of the inter-proton distance.}           
\label{f5}
\end{figure*}
\begin{table}
\caption{\label{t01} The equilibrium distance $R_{0}$, the equilibrium inverse size of the orbital $\alpha_{0}$, and the ground-state energy 
                     $E_{T}^{\left(4\right)}\left(R_{0}\right)$ for different values of $\gamma$.}
\begin{ruledtabular}
\begin{tabular}{|c||c|c|c|}
                        &                         &                                   &                               \\
 $\gamma$ [Ry]          & $R_{0}$ [${\rm a_{0}}$] & $\alpha_{0}$ [${\rm a^{-1}_{0}}$] & $E_{T}^{\left(4\right)}\left(R_{0}\right)$ [Ry] \\
                        &                         &                                   &                               \\
\hline
                        &                         &                                   &                               \\ 
  {\bf 0}               & {\bf 1.41968 }          & {\bf 1.199206}                    & {\bf -2.323011}               \\ 
   0.1                  & 1.413598                & 1.202479                          & -2.314919                     \\ 
   0.2                  & 1.396223                & 1.211990                          & -2.290874                     \\  
   0.3                  & 1.369845                & 1.22690                           & -2.251536                     \\  
   0.4                  & 1.33742                 & 1.24609                           & -2.19787                      \\  
   0.5                  & 1.301859                & 1.268341                          & -2.131022                     \\ 
   {\bf $\gamma_{\mathcal{PT}}=$0.520873}   &  {\bf 1.294281} & {\bf 1.273265} & {\bf -2.115516}                      \\ 
   0.6                                      &  1.265651       & 1.292526                          &  -2.052185        \\   
   {\bf $\gamma_{MS}=$0.659374}             &  {\bf 1.244701} &{\bf 1.307372}                     &  {\bf -2.000188}  \\  
   0.7                                      &  1.230858       & 1.317603                          &  -1.962537        \\
   0.8                                      &  1.199459       & 1.342514                          &  -1.863195        \\
   0.9                                      &  1.174508       & 1.365733                          &  -1.755232        \\
   1                                        &  1.168653       & 1.38188                           &  -1.639820        \\      
   {\bf $\gamma_{D}=$1.024638}              &  {\bf 1.196587} & {\bf 1.374634}                    &   {\bf -1.610491} \\ 
\end{tabular}
\end{ruledtabular}
\end{table}

Although the $\mathcal{PT}$ symmetry breaking does not manifest physically, the interaction of the hydrogen molecule with the environment can change its physical state. This fact is connected with the dependence of the total energy on the $\gamma$ parameter. 
In the figure \fig{f2}, we presented the total energy values ($E_{T}^{\left(j\right)}=E_{p}+E_{j}$) of the isolated hydrogen molecule, 
and the influence of the $\gamma$ parameter on the ground state energy $E_{T}^{\left(4\right)}$. 
One can see that with the increase of $\gamma$, we observe the increase of the minimum energy value $E_{T}^{\left(4\right)}\left(R_{0}\right)$, wherein 
the molecule is in the stable state. Above $\gamma_{MS}=0.659374$~Ry the hydrogen molecule can exist only in the metastable state: 
$E_{T}^{\left(4\right)}\left(R^{\left(MS\right)}_{0}\right)>E_{T}^{\left(4\right)}\left(R\rightarrow +\infty\right)=2$~Ry, 
where $R^{\left(MS\right)}_{0}=1.244701$~{$\rm a_{0}$}. After exceeding $\gamma_{D}=1.024638$~Ry, which corresponds to 
$R^{\left(D\right)}_{0}=1.196587$~{$\rm a_ {0}$}, the molecule breaks down. At the insert (a), we plotted the dependence of the hydrogen dissociation energy ($E_{\rm D}=2{\rm Ry}-E_{T}^{\left(4\right)}$) on the value of the $\gamma$ parameter. The insert (b) shows the influence of 
the $\gamma$ parameter on the equilibrium distance $R_{0}$. In the figures \fig{f3}~(a-c), we have traced the change of the distribution 
of electron charge for the stable case with $\gamma=0$ ($R_{0}=1.41968$~$ {\rm a_{0}}$), for the metastable case ($R^{\left(MS\right)}_{0}$), and 
at the dissociation point ($R^{\left(D\right)}_{0}$). The density of electron charge was calculated based on the formula: 
$\rho({\bf r})=\sum_{j}|\Phi_j\left({\bf r}\right)|^{2}$. 

The determination of the function $E_{T}\left(R\right)$ for the given parameter $\gamma$ 
allows to trace the influence of the environment on the vibrational energy. In the simplest approach (the harmonic approximation), the potential can be calculated as follows: 
$V_{\rm H}\left(R\right)=E_{T}^{\left(4\right)}\left(R_{0}\right)+\frac{1}{2}k_{\rm H}\left(R-R_ {0}\right)^{2}$, where: 
$k_{\rm H}=\left[d^{2}E_{T}^{\left(4\right)}\left(R\right)/dR^{2}\right]_{R=R_{0}}$. The quantum oscillator's energy has the form: 
\begin{equation}
\label{r01-III}
E^{\rm H}_{o}=\omega^{\rm H}_{0}\left(n+1/2\right).
\end{equation}
The symbol $n$ indexes the energy level: $n=0,1,2, ...$. Additionally, $\omega^{\rm H}_{0}=\sqrt{k_{\rm H}/m'}$, with $m'$ is the reduced mass 
of the proton: $m'=m_{p}/2=918.076336$ ($m_{p}$ is the proton mass). The more advanced approach is based on the Morse potential:
$V_{\rm Mo}\left(R\right)=E_{T}^{\left(4\right)}\left(R_{0}\right)+E_{\rm D}\left[1-\exp\left(-\alpha_{\rm Mo}\left(R-R_{0}\right)\right)\right]^{2}$, 
where $\alpha_{\rm Mo}$ means measure of the curvature of the potential about its minimum. The force constants, $k_{\rm Mo}$ should be calculated 
based on the formula: $k_{\rm Mo}=\left[d^{2}V_{\rm Mo}\left(R\right)/dR^{2}\right]_{R=R_{0}}$. The Morse energy is given by: 
$\omega^{\rm Mo}_{0}=\sqrt{k_{\rm Mo}/m'}$ (see table \tab{tB2}). The energy formula possesses the more complex form than in the harmonic case:
\begin{equation}
\label{r02-III}
E^{\rm Mo}_{o}=\omega^{\rm Mo}_{0}\left(n+1/2\right)+\left((\omega^{\rm Mo}_{0})^{2}/4E_{D}\right)\left(n+1/2\right)^{2}.
\end{equation}
In the figure \fig{f4}~(a), we plotted the dependency of the energies $\omega^{\rm H}_{0}$ and $\omega^{\rm Mo}_{0}$ on the value of the 
$\gamma$ parameter. There is a clear difference in the course of the functions under consideration. It results from the method of approximating 
of the exact dependence of the total energy on the inter-proton distance (see figure \fig{f4}~(b)). We notice that the anharmonic approximation 
can be used only for the $\gamma$ values smaller than $\gamma_{MS}$, for higher values, the Morse curve incorrectly parameterizes the ground state energy function $E_{T}^{\left(4\right)}\left(R\right)$.

The rotational energy of the hydrogen molecule should be calculated on the basis of the expression:
\begin{equation}
\label{r03-III}
E_{r}=B_{0}l\left(l+1\right),
\end{equation}
where: $B_{0}=1/m' R^{2}_{0}$ and $l=0,1,2,...$. The influence of the $\gamma$ parameter on the rotational energy value 
has been presented in the figure \fig{f4}~(c). From the physical point of view, the increase of the energy $B_{0}$ results 
from the decrease of the equilibrium distance $R_{0}$, which we observe when the $\gamma$ parameter grows (see the figure \fig{f2} - inset (b)).

Having the explicit dependence of the $\hat{\mathcal{H}}_{e\gamma}$ parameters on $R$ (see the appendix A), 
we computed the electron-phonon couplings based on the formula: $g_{x}=dx/dR$, where $x\in\{\varepsilon,t,U,K,J,V\}$. 
We plotted the obtained results in the figure \fig{f5}. It is easy to see that the absolute values considered functions at $R_{0}$ increases 
as the $\gamma$ parameter increases. The greatest physical importance possess the couplings associated with the $\varepsilon$, $t$, $U$, and $K$ parameters. The other two quantities can be omitted in the considerations. Note the relatively high values of the $g_{U}$ and $g_{K}$ functions. 
The obtained result is due to the fact that the electrons in the hydrogen molecule form the strongly correlated system.

\subsection{The dynamic instability of the hydrogen molecule}

\begin{figure*}
\includegraphics[width=0.49\columnwidth]{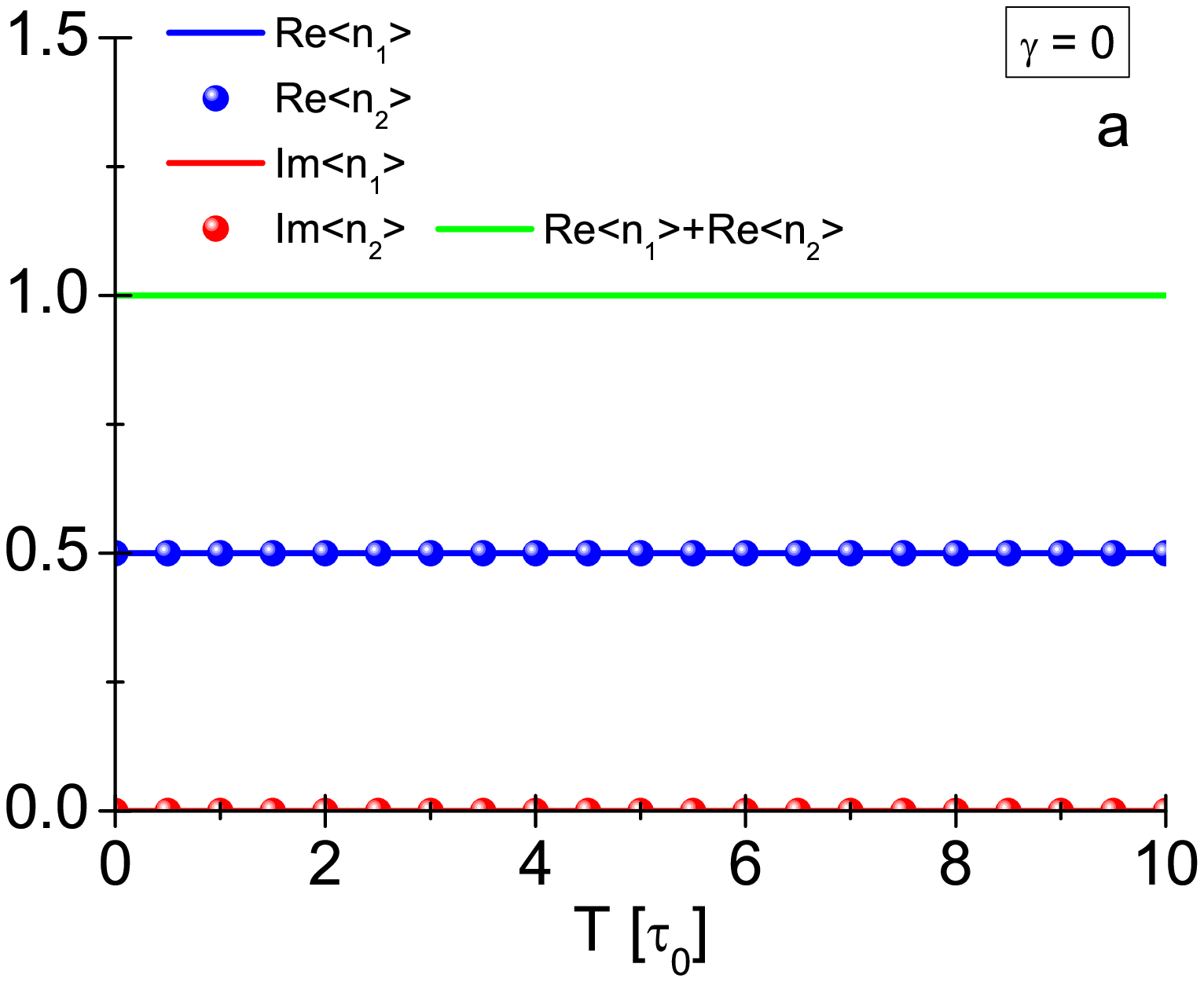}
\includegraphics[width=0.49\columnwidth]{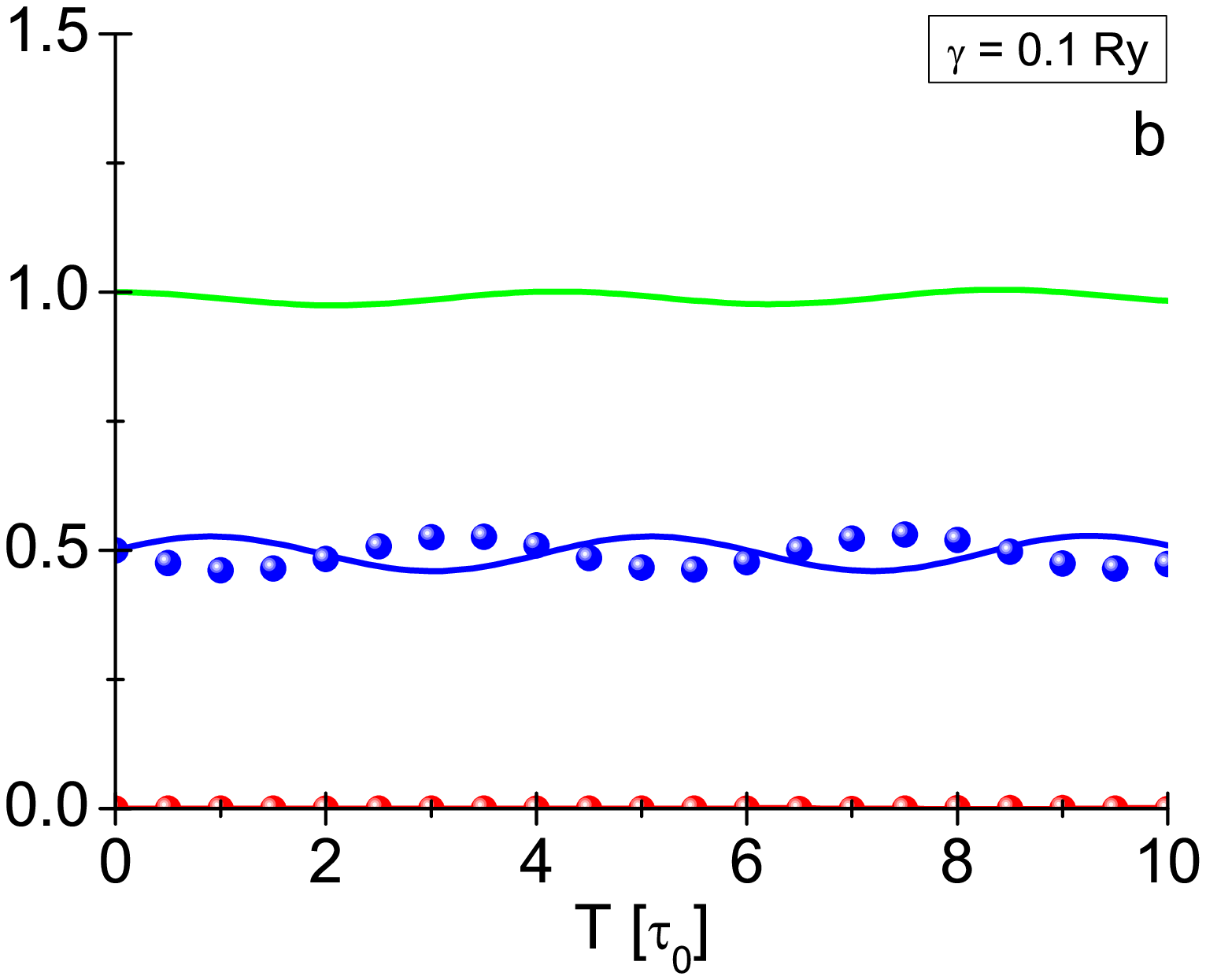}
\includegraphics[width=0.49\columnwidth]{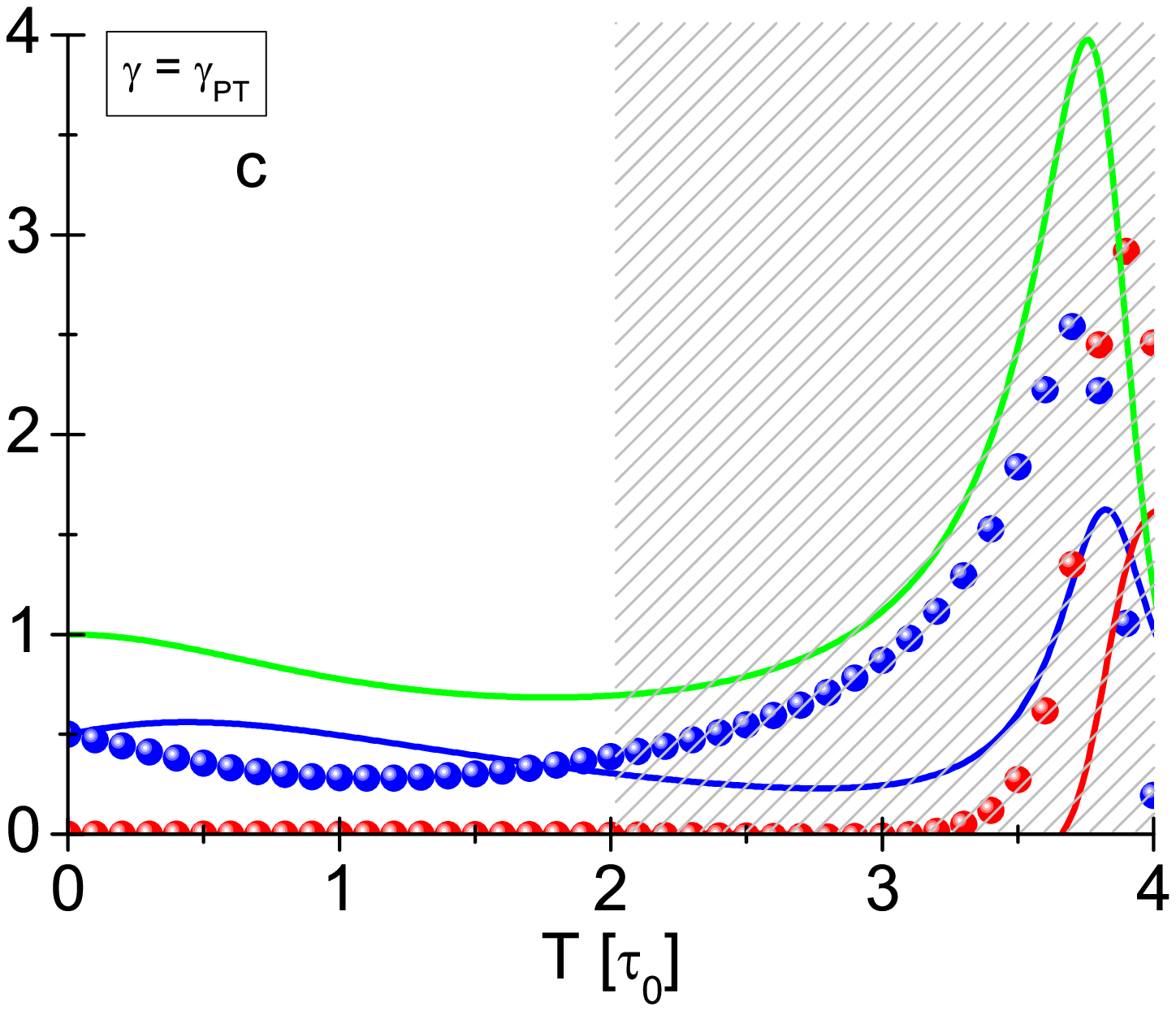}
\includegraphics[width=0.49\columnwidth]{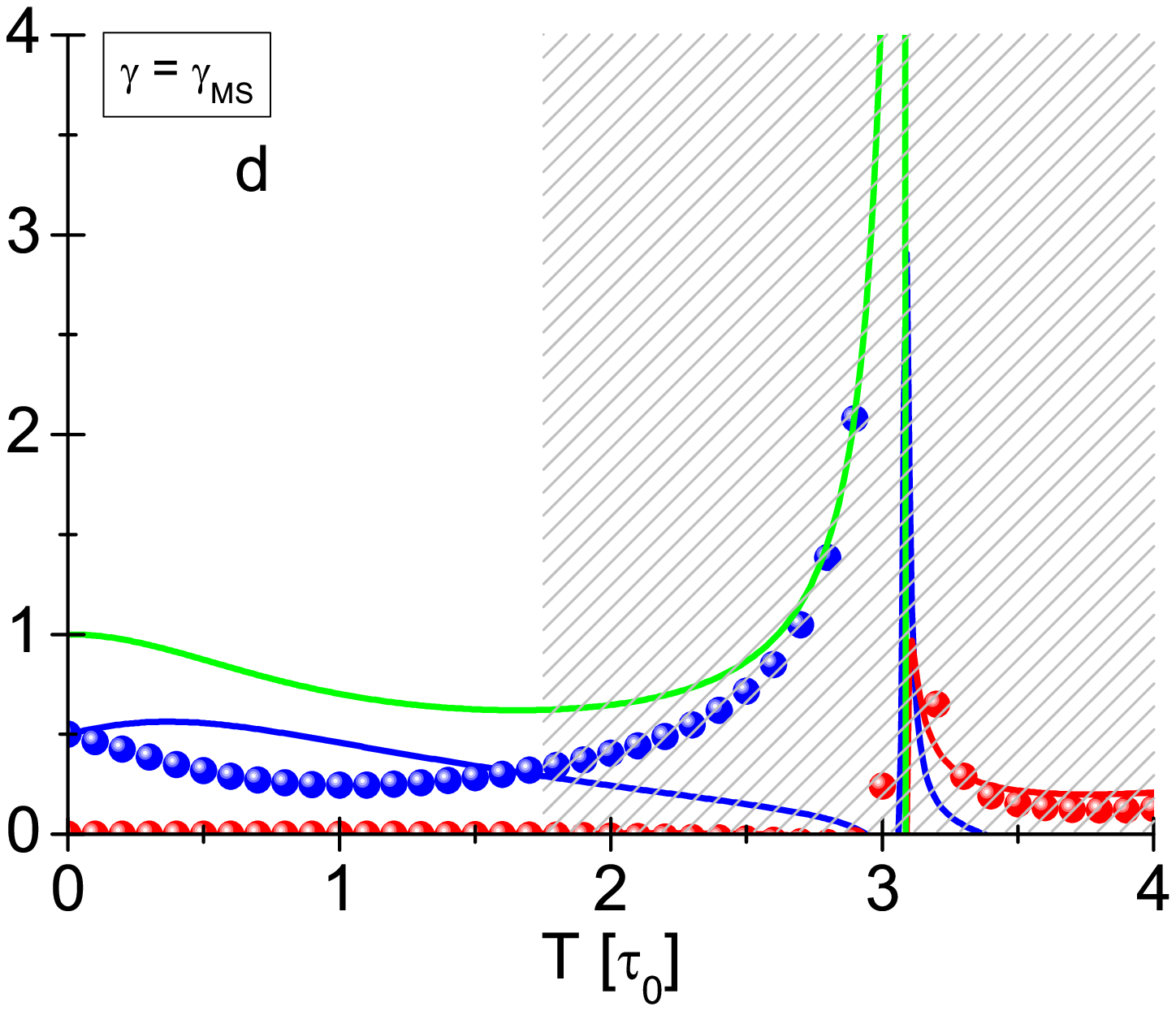}
\includegraphics[width=0.49\columnwidth]{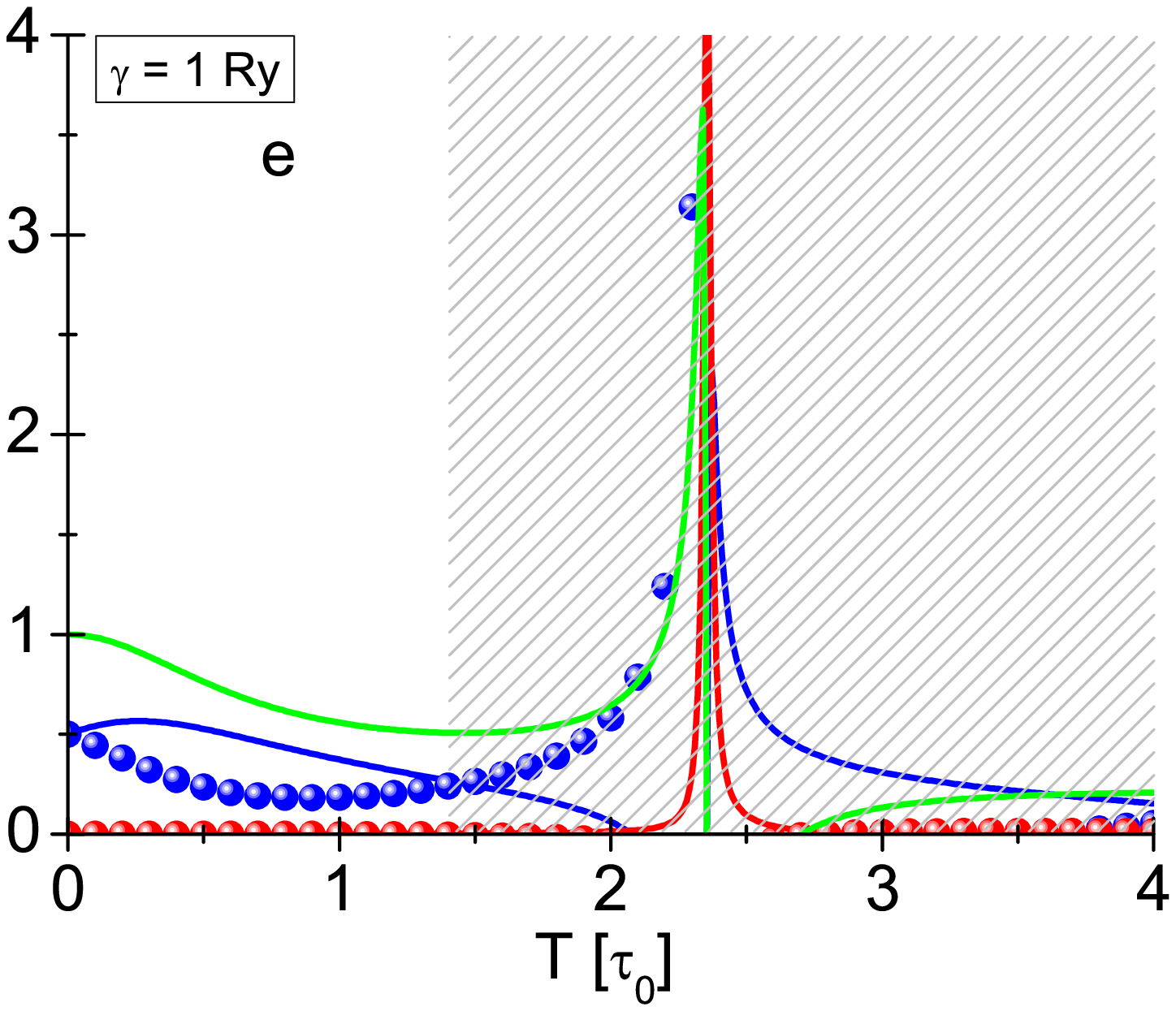}
\includegraphics[width=0.49\columnwidth]{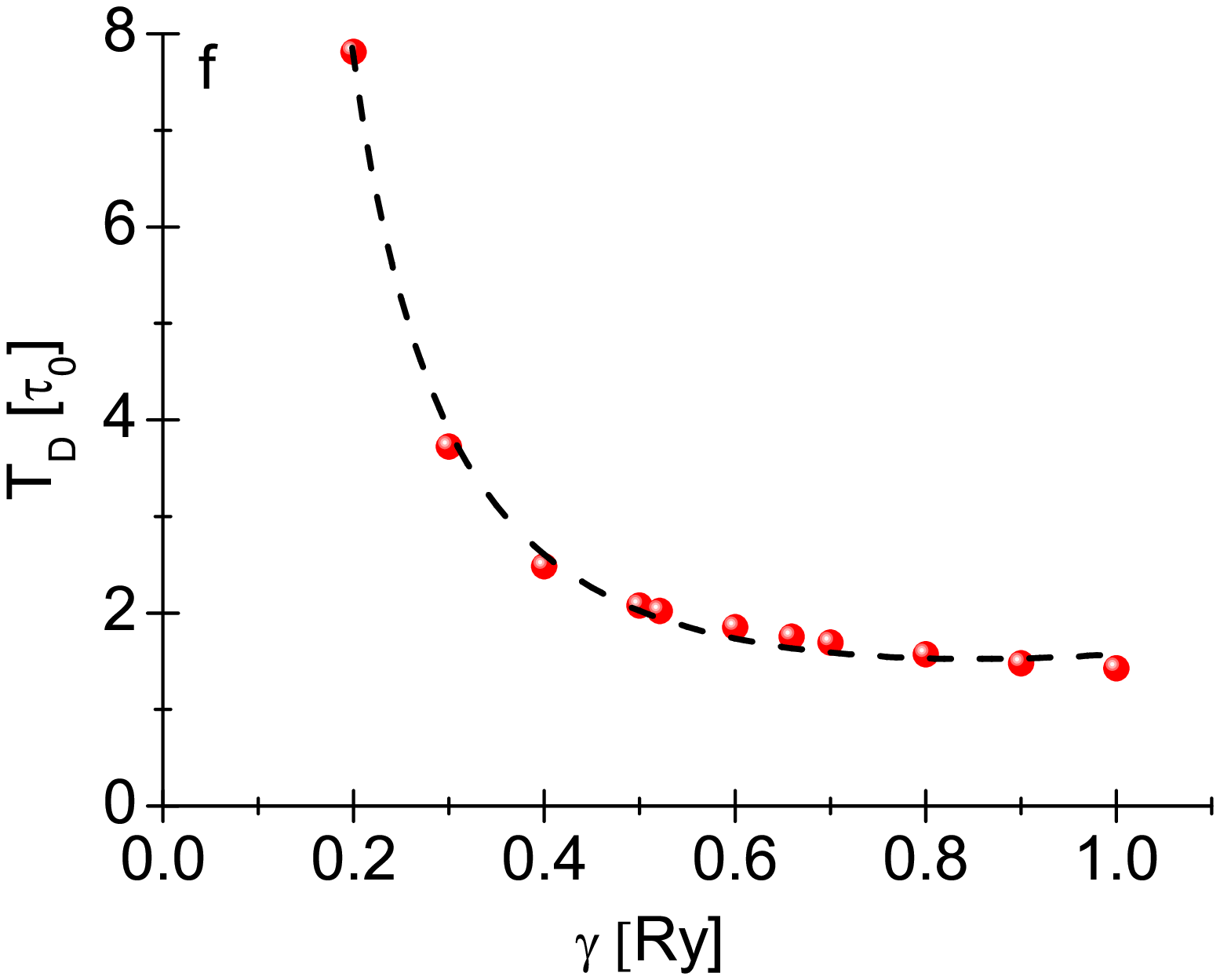}
\caption{(a)-(e) The time evolution of $\left<n_{1\uparrow}\right>$ and $\left<n_{2\uparrow}\right>$ respectively 
         for $\gamma$ equal to: $0$, $0.1$~Ry, $\gamma_{\mathcal {PT}}$, $\gamma_{MS}$, and $1$~Ry. 
         The hatched areas correspond to the value greater than $T_{D}$.
         Figure (f) shows the form of $T_{D}\left(\gamma\right)$ function.
         The dashed curve was obtained from the formula: $T_{D}=\frac{a}{\gamma^{b}}\exp\left(c\gamma\right)$, 
         where $a=0.095689$, $b=2.38294$, and $c=2.80013$.}           
\label{f6}
\end{figure*}

The basic observable of the electronic subsystem is the occupation number $\left<\hat{n}_{j\sigma}\right>$, 
where the symbol $\left<...\right> $ means the expectation value. In the hermitian case ($\gamma=0$) 
the dynamics of $\left<\hat{n}_{j\sigma}\right>$ can be analyzed using the Heisenberg equation: 
\begin{eqnarray}
\label{r01-IIIb}
i\frac{d\left<\hat{n}_{j\sigma}\right>}{dT}=\left<\left[\hat{n}_{j\sigma},\hat{\mathcal{H}}^{MF}_{e\gamma}\right]_{-}\right>,
\end{eqnarray}
where $\hat{\mathcal{H}}^{MF}_{e\gamma}$ is the mean field Hamiltonian:
\begin{eqnarray}
\label{r02-IIIb}
\hat{\mathcal{H}}^{MF}_{e\gamma}=\sum_{j\sigma}\varepsilon_{j\sigma}\left(\gamma\right)\hat{n}_{j\sigma}+\sum_{j\sigma}t_{j\sigma}
\hat{n}_{j\overline{j}\sigma}+\sum_{j\sigma}J_{j\sigma}\hat{n}_{j\sigma-\sigma}+\sum_{j}\left(P_{j}\hat{\Delta}^{\dagger}_{j}+P^{\star}_{j}\hat{\Delta}_{j}\right).
\end{eqnarray}
The Hamiltonian parameters have been defined by the expressions:
\begin{eqnarray}
\label{r03-IIIb}
\varepsilon_{j\sigma}\left(\gamma\right)&=&\varepsilon+U\left<\hat{n}_{j-\sigma}\right>
+K\sum_{\sigma'}\left<\hat{n}_{\overline{j}\sigma'}\right>
-J\left<\hat{n}_{\overline{j}\sigma}\right>+V\left[\left<\hat{n}_{j\overline{j}-\sigma}\right>+\left<\hat{n}_{\overline{j}j-\sigma}\right>\right]
-\left(-1\right)^{j}i\gamma,
\\ \nonumber
t_{j\sigma}&=&t+V\left[\left<\hat{n}_{j-\sigma}\right>+\left<\hat{n}_{\overline{j}-\sigma}\right>\right],\\ \nonumber
J_{j\sigma}&=&-J\left<\hat{n}_{\overline{j}-\sigma\sigma}\right>, \\ \nonumber
P_{j}&=&J\left<\hat{\Delta}_{\overline{j}}\right>.
\end{eqnarray}
The new symbols have the following meanings:
\begin{equation}
\label{r04-IIIb}
\overline{j}=
\left\{ \begin{array}{l} 1 \hspace{5mm} {\rm for} \hspace{5mm} j=2\\
                         2 \hspace{5mm} {\rm for} \hspace{5mm} j=1, \end{array} \right.
\end{equation}
$\hat{n}_{j\overline{j}\sigma}=\hat{c}^{\dagger}_{j\sigma}\hat{c}_{\overline{j}\sigma}$, 
$\hat{n}_{j\sigma-\sigma}=\hat{c}^{\dagger}_{j\sigma}\hat{c}_{j-\sigma}$, and $\hat{\Delta}_{j}=\hat{c}_{j\downarrow}\hat{c}_{j\uparrow}$. 

We emphasize that for $\hat{\mathcal{H}}_{e\gamma}$ the required operator's calculations are not feasible due to their size.
The mean-field approximation transforms the operator $\hat{\mathcal{H}}_{e\gamma}$ into the Hamiltonian, in which the energy of the molecular state 
and the hopping integral explicitly depend on the proton index $j$ and the spin $\sigma$. In addition, the Hamiltonian have the part that models 
the reversal of the spin due to the exchange interaction $J$. It is also worth paying attention to the quantity of $\hat{\Delta}_{j}$, 
which has the formal structure of the Cooper's annihilation operator in the real space. 
This analogy is not complete, because the Hamiltonian $\hat{\mathcal{H}}^{MF}_{e\gamma}$ containing $\hat{\Delta}_{j}$ and $\hat{\Delta}^{\dagger}_{j}$ 
does not correspond to BCS pairing operator \cite{Bardeen1957A, Bardeen1957B, Szczesniak2012A} 
(the integral of the exchange $J_{0}$ has the positive value instead of negative - see the table \tab{tA1}).  
   
After performing the required operator calculations, we get the set of sixteen first-order differential equations, which is explicitly written 
in the appendix \ref{dodD}. In the non-hermitian case ($\gamma\neq 0$), determining the time dependence of the electron observables is the more subtle issue \cite{Dattoli1990A, Graefe2008A, Sergi2013A}. First of all, one must define the operators: 
$\hat{\mathcal{H}}^{MF}_{e\gamma\pm}=\frac{1}{2}\left(\hat{\mathcal{H}}^{MF}_{e\gamma}\pm\hat{\mathcal{H}}^{MF\dagger}_{e\gamma}\right)$, where 
$\hat{\mathcal{H}}^{MF}_{e\gamma\pm}=\pm\hat{\mathcal{H}}^{MF\dagger}_{e\gamma\pm}$. Then we use the generalized form of the Heisenberg equation:  

\begin{eqnarray}
\label{r05-IIIb}
i\frac{d\left<\hat{n}_{j\sigma}\right>}{dT}=
\left<\left[\hat{n}_{j\sigma},\hat{\mathcal{H}}^{MF}_{e\gamma+}\right]_{-}\right>
+\left<\left[\hat{n}_{j\sigma},\hat{\mathcal{H}}^{MF}_{e\gamma-}\right]_{+}\right>
-2\left<\hat{n}_{j\sigma}\right>\left<\hat{\mathcal{H}}^{MF}_{e\gamma-}\right>.
\end{eqnarray}
Tedious but not difficult operator calculations lead to the complex system of the differential equations, which form is presented 
in the appendix \ref{dodE}.

In the figures \fig{f6}~(a)-(e), we plotted the time dependence of the observables $\left<n_{1\uparrow}\right>$, 
$\left<n_{2\uparrow}\right>$ for the selected values of the $\gamma$ parameter. As expected, one can see that in the hermitian 
case the system is in the stable state, which manifests itself time constancy of the expectation value. 
In the non-hermitian case, the weak interaction of the hydrogen molecule with the environment ($\gamma=0.1$) causes oscillatory changes 
in the time of the discussed quantities. However, this is not time-stable state of the system, because from the specific 
moment $T_{D}$ observables 
$\left<n_{1\uparrow}\right>$, $\left<n_{2\uparrow}\right>$ accept complex values. 
From the physical point of view, the time $T_{D}$ should be interpreted as the moment in which the system dissociates. 
It is easy to show that as the $\gamma$ parameter increases, the oscillations of the expectation values disappear 
and the value $T_{D}$ decreases very clearly (see the figure \fig{f6}~(f)). The results obtained mean that any weak interaction of the hydrogen molecule with the environment modeled in the BGL scheme leads to the finite life time of the molecule.

\section{Summary and discussion of the results}

The obtained results show that the BGL type interaction of the hydrogen molecule with the environment leads to its disintegration. 
This result is due to the dynamic instability of the electronic subsystem. Note that the dynamic instability of the molecule overlaps the static instability for high values of $\gamma$ parameter. We have shown that the increase in the value of $\gamma$ strongly reduces 
the dissociation energy of the molecule. Above $\gamma_{MS}=0.659374$~Ry, the molecule is in the metastable state, decaying definitively 
for  $\gamma_{D}>1.024638$~Ry. 

An additional effect, that we observed for $\gamma$ higher than $\gamma_{\mathcal{PT}}=0.520873$~Ry, is the $\mathcal{PT}$ symmetry breaking 
of the electronic Hamiltonian $\hat{\mathcal{H}}_{e\gamma}$. As a result, the two highest energies of the electron state assume complex values 
and the number of available electronic states of the molecule is reduced to four. This effect does not affect the stability of the tested system.   
Additionally, the $\mathcal{PT}$ symmetry breaking does not change the values of the integrals of the electronic Hamiltonian, 
the phonon or rotational properties of the hydrogen molecules and the electron-phonon interaction constants. 
The dynamics of the electronic subsystem is also independent on the breaking of the $\mathcal{PT}$ symmetry of $\hat{\mathcal{H}}_{e\gamma}$.
 
\appendix

\newpage
\section{\label{dodA} The eigenvalues of the electronic Hamiltonian of the hydrogen molecule interacting with the environment}

The Hamiltonian $\hat{\mathcal{H}}_{e\gamma}$ should be written in the matrix form:
\begin{equation}
\label{r01-A}
\left(
\begin{array}{cccccc}
h^{+}_{1}               &  0                    & t+V             & 0                     & J                       &  t+V            \\
0                       &  h_{2}                & 0               & 0                     & 0                       &  0              \\
t+V                     &  0                    & 2\varepsilon+K  & 0                     & t+V                     &  -J             \\
0                       &  0                    & 0               & h_{2}                 & 0                       &  0              \\
J                       &  0                    & t+V             & 0                     & h^{-}_{1}               &  t+V            \\
t+V                     &  0                    & -J              & 0                     & t+V                     & 2\varepsilon+K    
\end{array}
\right).
\end{equation}
where: $h^{\pm}_{1}=2\varepsilon+U\pm 2i\gamma$, and $h_{2}=2\varepsilon+K-J$. By using the operator (\ref{r01-A}) there was brought out the preliminary formulas for the eigenvalues, which has the form as follows:
\begin{equation}
\label{r02-A}
E_1=-J+K+2\varepsilon,
\end{equation}
\begin{equation}
\label{r03-A}
E_2=-J+K+2\varepsilon,
\end{equation}
\begin{equation}
\label{r04-A}
E_3=J+K+2\varepsilon,
\end{equation}
\begin{widetext}
\begin{eqnarray}
\label{r05-A}
E_4
&=&
\frac{1}{3}(-J+K+2U+[-4J^2+2J(K-U)-
(K-U)^2+12(-(t+V)^2+\gamma^2)]/\\ \nonumber
& &
[-8J^3+6J^2(K-U)+
3J((K-U)^2+12(-(t+V)^2+\gamma^2))-(K-U)((K-U)^2\\ \nonumber
&+&
18((t+V)^2+2\gamma^2))
+\frac{1}{2}\sqrt{A-B}]^{\frac{1}{3}}-
[-8J^3+6J^2(K-U)+3J((K-U)^2\\ \nonumber
&+&
12(-(t+V)^2+\gamma^2))-(K-U)((K-U)^2+18((t+V)^2 
+
2\gamma^2))+\frac{1}{2}\sqrt{A-B}]^{\frac{1}{3}}+6\epsilon),
\end{eqnarray}
\begin{eqnarray}
\label{r06-A}
E_5
&=&
\frac{1}{12}([2(1+i\sqrt{3})(4J^2+(K-U)^2+2J(-K+U)+12(t+V-\gamma)(t+V+\gamma))]/\\ \nonumber
& &
[-8J^3+6J^2(K-U)
+
3J((K-U)^2+12(-(t+V)^2+\gamma^2))-(K-U)((K-U)^2\\ \nonumber
&+&
18((t+V)^2+2\gamma^2))
+\frac{1}{2}\sqrt{A-B}]^{\frac{1}{3}}
+
2(1-i\sqrt{3})[-8J^3+6J^2(K-U)+3J((K-U)^2\\ \nonumber
&+&
12(-(t+V)^2+\gamma^2))-(K-U)((K-U)^2+18((t+V)^2
+
2\gamma^2))+\frac{1}{2}\sqrt{A-B}]^{\frac{1}{3}}\\ \nonumber
&+&
4(-J+K+2U+6\varepsilon)),
\end{eqnarray}
\begin{eqnarray}
\label{r07-A}
E_6
&=&
\frac{1}{12}([2(1-i\sqrt{3})(4J^2+(K-U)^2+2J(-K+U)+12(t+V-\gamma)(t+V+\gamma))]/\\ \nonumber
& &
[-8J^3+6J^2(K-U)
+
3J((K-U)^2+12(-(t+V)^2+\gamma^2))-(K-U)((K-U)^2\\ \nonumber
&+&
18((t+V)^2+2\gamma^2))
+\frac{1}{2}\sqrt{A-B}]^{\frac{1}{3}}
+
2(1+i\sqrt{3})[-8J^3+6J^2(K-U)+3J((K-U)^2\\ \nonumber
&+&
12(-(t+V)^2+\gamma^2))-(K-U)((K-U)^2+18((t+V)^2
+
2\gamma^2))+\frac{1}{2}\sqrt{A-B}]^{\frac{1}{3}}\\ \nonumber
&+&
4(-J+K+2U+6\varepsilon)),
\end{eqnarray}
\end{widetext}
wherein:
\begin{eqnarray}
\label{r08-A}
A=4[(2J+K-U)[(J-K+U)(4J-K+U)+18(t+V)^2]-36(J-K+U)\gamma^2]^2,
\end{eqnarray}
and
\begin{eqnarray}
\label{r09-A}
B=4[4J^2+(K-U)^2+2J(-K+U)+12(t+V-\gamma)(t+V+\gamma)]^3.
\end{eqnarray}

\begin{figure*}
\includegraphics[width=\columnwidth]{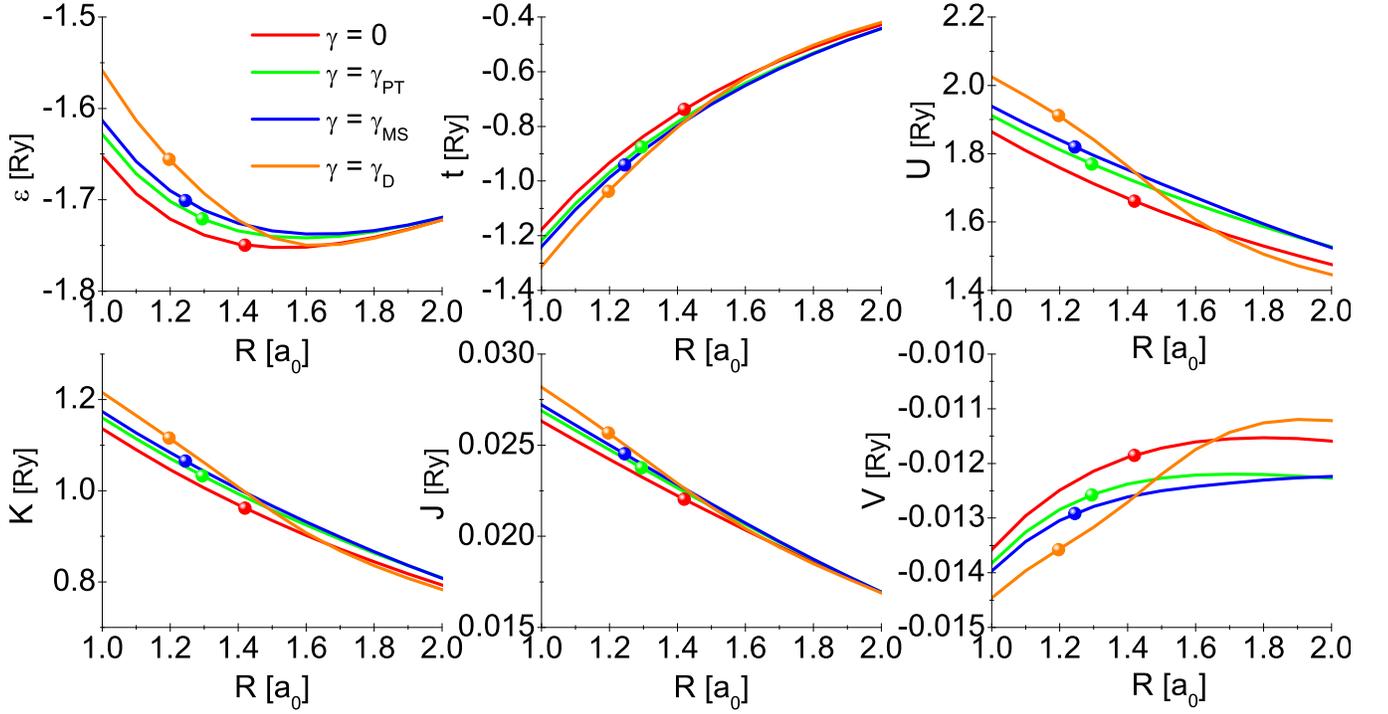}
\caption{The integral of the Hamiltonian $\hat{\mathcal{H}}_{e\gamma}$ as a function of the inter-proton distance for the selected
         values of the parameter modeling the interaction of the molecule with the environment. The balls placed on the curves point 
         to the equilibrium value of the inter-proton distance $R_{0}$.}           
\label{fA1}
\end{figure*}

\begin{figure*}
\includegraphics[width=0.5\columnwidth]{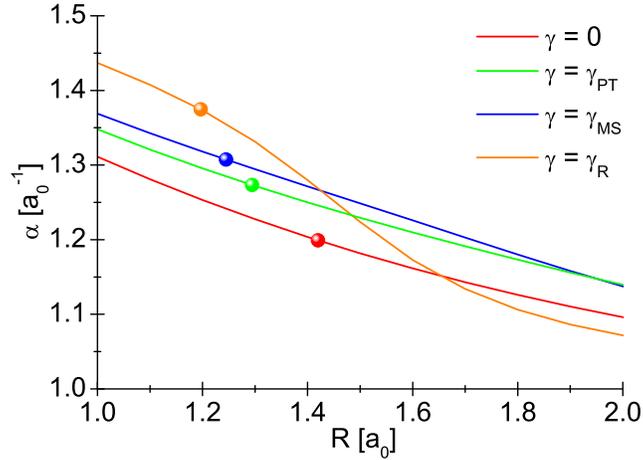}
\caption{The variation parameter $\alpha$ as a function of the proton distance for selected $\gamma$ values.}           
\label{fA2}
\end{figure*}

An attentive reader will notice that the energies $\varepsilon$, $t$, $U$, etc. are explicit functions of the inter-proton distance $R$ and the parameter 
$\alpha$. In the figure \fig{fA1}, we plotted the discussed values of the energies as the function of $R$ and $\gamma$. 
Additionally, in table \tab{tA1} we give the equilibrium values of the $\Hat{\mathcal{H}}_{e\gamma}$ parameters. 
The explicit dependence of the variational parameter $\alpha$ on the distance $R$ is shown in the figure \fig{fA2}.

\begin{table*}
\caption{\label{tA1} The values of the Hubbard Hamiltonian integrals calculated for the equilibrium distance 
                     of the hydrogen molecule. The selected values $\gamma$ have been taken into account. 
                    }
\begin{ruledtabular}
\begin{tabular}{|c||c|c|c|c|c|c|}
                       &                     &                &                 &                 &                &                  \\
$\gamma$ [Ry]          & $\varepsilon_{0}$ [Ry]  & $t_{0}$ [Ry]  & $U_{0}$ [Ry] & $K_{0}$ [Ry] & $J_{0}$ [Ry] & $V_{0}$ [Ry]          \\
                       &                     &                &                 &                 &                &                  \\
\hline
                       &                     &                &                 &                 &                &                  \\
{\bf 0}        & {\bf -1.749493}  & {\bf -0.737679}& {\bf 1.661254}  &  {\bf 0.962045} & {\bf 0.022040} &  {\bf -0.011851} \\ 
         0.1           & -1.74866                   & -0.743562              &  1.66607              &  0.965198              & 0.022117              &  -0.0118825               \\
         0.2           & -1.74599                   & -0.760758              &  1.68006              &  0.974349              & 0.0223398             &  -0.0119744               \\
         0.3           & -1.74114                   & -0.788025              &  1.70198              &  0.988664              & 0.0226876             &  -0.0121193               \\
         0.4           & -1.73366                   & -0.823632              &  1.73016              &  1.00702              & 0.0231322              &  -0.0123074               \\
         0.5           & -1.72329                   & -0.865608              &  1.76279              &  1.02821              & 0.0236434              &  -0.0125275               \\
{\bf $\gamma_{\mathcal{PT}}=$0.520873} & {\bf -1.72075} & {\bf -0.874986} &  {\bf 1.770000} & {\bf  1.03288 } & {\bf 0.0237558} & {\bf -0.0125764} \\
         0.6                           & -1.70997       & -0.911931       &  1.79817        &  1.05107        & 0.0241927       &  -0.0127687      \\
{\bf $\gamma_{MS}=$0.659374}           & {\bf -1.70076} & {\bf  -0.940627}&  {\bf  1.81984} & {\bf 1.064996}  & {\bf 0.0245259} &  {\bf -0.0129176}\\
         0.7                           & -1.69397       & -0.960458       &  1.83473        &  1.07453        & 0.0247533       &  -0.0130205      \\
         0.8                           & -1.67604       & -1.00854        &  1.87081        &  1.09738        & 0.0252963       &  -0.0132714      \\
         0.9                           & -1.65787       & -1.0515         &  1.90394        &  1.11776        & 0.0257748       &  -0.0135041      \\
         1.0                           & -1.6462        & -1.07108        &  1.92509        &  1.1285         & 0.0260069       &  -0.0136584      \\          
{\bf $\gamma_{D}=$1.024638}            & {\bf -1.65569} & {\bf -1.03796}  &  {\bf  1.911296}&  {\bf  1.11581} & {\bf 0.0256683} &{\bf -0.0135781}  \\     
                                       &                &                 &                 &                 &                &                   \\      
\end{tabular}
\end{ruledtabular}
\end{table*}

\newpage
\section{\label{dodB} The equilibrium values of phonon energy, rotational energy and the electron-phonon coupling function}

In the table \tab{tB1} and \tab{tB2}, we collected the equilibrium values of the phonon parameters for selected $\gamma$. 
The table \tab{tB3} presents the equilibrium values of the electron-phonon coupling functions.
\begin{table*}
\caption{\label{tB1} The harmonic potential parameter $k_{\rm H}$ and the quantum energy for different values of $\gamma$.}
\begin{ruledtabular}
\begin{tabular}{|c||c|c|}
                               &      &    \\
$\gamma$ [Ry]         & $k_{\rm H}$  [${\rm Ry/a^{2}_{0}}$] & $\omega^{\rm H}_{0}$ [Ry] \\
                               &      &    \\ 
\hline
                                       &                  &                           \\ 

{\bf 0}                                &  {\bf 0.691719}  &{\bf   0.027449}  \\ 
0.1                                    &       0.379254   &       0.027886   \\ 
0.2                                    &       0.387102   &       0.028463   \\ 
0.3                                    &       0.402162   &       0.029570   \\ 
0.4                                    &       0.412453   &       0.0303274  \\ 
0.5                                    &       0.427453   &       0.0314304  \\ 
{\bf $\gamma_{\mathcal{PT}}=$0.520873} &  {\bf 0.919309}  &  {\bf 0.031644}  \\ 
0.6                                    &       0.439769   &       0.0323359  \\ 
{\bf $\gamma_{\mathcal{MS}}=$0.659374} &  {\bf 0.980341}  &  {\bf 0.0326775} \\ 
0.7                                    &       0.0327805  &       0.445815   \\ 
0.8                                    &       0.440037   &       0.0323556  \\ 
0.9                                    &       0.409835   &       0.030135   \\ 
1                                      &       0.29284    &       0.0215324  \\ 
{\bf $\gamma_{\mathcal{D}}=$1.024638}  &  {\bf 0.0050384} & {\bf 0.00234265} \\ 
                                       &                  &                  \\       
\end{tabular}
\end{ruledtabular}
\end{table*}
\begin{table*}
\caption{\label{tB2} The Morse potential parameters for different values of $\gamma$.}
\begin{ruledtabular}
\begin{tabular}{|c||c|c|}
                                        &                   &                                       \\ 
 $\gamma$ [Ry]                          & $E_{\rm D}$  [Ry] & $\alpha_{\rm M}$ [${\rm a^{-1}_{0}}$] \\                        
                                        &                   &                                       \\ 
\hline
                                        &                   &                                       \\ 
 {\bf 0}                                & {\bf 0.323007}    & {\bf 1.441564}                        \\ 
 0.1                                    &      0.314916     &      1.19386                          \\ 
 0.2                                    &      0.290867     &      1.24564                          \\
 0.3                                    &      0.251537     &      1.33815                          \\
 0.4                                    &      0.19483      &      1.48597                          \\
 0.5                                    &      0.123658     &      1.76235                          \\
 {\bf $\gamma_{\mathcal{PT}}=$0.520873} & {\bf 0.114146}    & {\bf 1.816998}                        \\
 0.6                                    &      0.062886     &      2.19766                          \\
 {\bf $\gamma_{\mathcal{MS}}=$0.659374 }& {\bf 0.0194576}   & {\bf 2.95769}                         \\
                                        &                   &                                       \\ 
\end{tabular}
\end{ruledtabular}
\end{table*}
\begin{table*}
\caption{\label{tB3} The values of the electron-ion coupling constants at the hydrogen-molecule equilibrium for different values of $\gamma$.}
\begin{ruledtabular}
\begin{tabular}{|c||c|c|c|c|c|c|}
                              &                         &                  &              &              &              &             \\ 
 $\gamma$ [Ry]       & $g_{\varepsilon_{0}}$ [${\rm Ry/a_{0}}$] & $g_{t_{0}}$ [${\rm Ry/a_{0}}$]
       &$g_{U_{0}}$ [${\rm Ry/a_{0}}$] & $g_{K_{0}}$ [${\rm Ry/a_{0}}$] & $g_{J_{0}}$ [${\rm Ry/a_{0}}$] & $g_{V_{0}}$ [${\rm Ry/a_{0}}$]\\
                              &                         &                  &              &              &              &             \\
\hline
                              &                &                &                 &                 &                 &                \\
       {\bf 0}                & {\bf 0.001744} & {\bf 0.609033} & {\bf -0.126289} & {\bf -0.236261} & {\bf -0.007502} & {\bf -0.000385}\\
0.1  & -0.000858724 & 0.615157 & -0.127093 & -0.237581 & -0.00752746 & -0.000362566 \\
0.2  & -0.00860958  & 0.633159 & -0.12944  & -0.241435 & -0.00760231 & -0.000297894 \\
0.3  & -0.0213324   & 0.662009 & -0.133151 &-0.247532  & -0.00772034 & -0.000194141 \\
0.4  & -0.0387245   & 0.70024  & -0.13798  & -0.255475 & -0.00787403 & -0.0000572511 \\
0.5  & -0.0603282   & 0.746108 & -0.143655 & -0.264823 & -0.00805519 & 0.000105421 \\
{\bf $\gamma_{\mathcal{PT}}=$ 0.520873 } & {\bf -0.0653133 }& {\bf 0.756473 }&{\bf  -0.144921 }&{\bf  -0.266911} &{\bf  -0.00809584 }&{\bf  0.000141774} \\
0.6  & -0.0854991 & 0.797725 & -0.149904 & -0.27515 & -0.00825739 & 0.000284209 \\
{\bf $\gamma_{\mathcal{MS}}=$ 0.659374}  &{\bf  -0.101762}&{\bf  0.830226 }&{\bf  -0.153775 }& {\bf -0.281576 }& {\bf -0.00838543 }& {\bf 0.000392951 }\\
0.7  & -0.113288 & 0.852925 & -0.156455 & -0.286042 & -0.00847612 & 0.000466248 \\
0.8  & -0.142132 & 0.908801 & -0.162993 & -0.297043 & -0.00870995 & 0.00063131  \\
0.9  & -0.168739 & 0.95991  & -0.168994 & -0.307449 & -0.00896312 & 0.000736959 \\
1    & -0.179745 & 0.984777 & -0.17245  & -0.314684 & -0.0092668  & 0.000605461 \\
{\bf $\gamma_{\mathcal{D}}=$ 1.024638 } &{\bf  -0.156139 }&{\bf 0.947351 } &{\bf  -0.169079 }&{\bf -0.311133 }&{\bf -0.00941164 } &{\bf  0.000204999 }\\
     &           &          &           &                 &                &     \\
\end{tabular}
\end{ruledtabular}
\end{table*}

\newpage
\section{\label{dodD} The set of the differential equations for electron observables ($\gamma=0$)}

The system of differential equations has the form:
\begin{eqnarray}
\label{r01-D}
i\frac{d\left<\hat{n}_{1\uparrow}\right>}{dT}=
t_{1\uparrow}\left<\hat{n}_{12\uparrow}\right>-t_{2\uparrow}\left<\hat{n}_{21\uparrow}\right>
+J_{1\uparrow}\left<\hat{n}_{1\uparrow\downarrow}\right>-J_{1\downarrow}\left<\hat{n}_{1\downarrow\uparrow}\right>+
P_{1}\left<\hat{\Delta}^{\dagger}_{1}\right>-P^{\star}_{1}\left<\hat{\Delta}_{1}\right>,
\end{eqnarray}
\begin{eqnarray}
\label{r02-D}
i\frac{d\left<\hat{n}_{1\downarrow}\right>}{dT}=
t_{1\downarrow}\left<\hat{n}_{12\downarrow}\right>-t_{2\downarrow}\left<\hat{n}_{21\downarrow}\right>
+J_{1\downarrow}\left<\hat{n}_{1\downarrow\uparrow}\right>-J_{1\uparrow}\left<\hat{n}_{1\uparrow\downarrow}\right>+
P_{1}\left<\hat{\Delta}^{\dagger}_{1}\right>-P^{\star}_{1}\left<\hat{\Delta}_{1}\right>,
\end{eqnarray}
\begin{eqnarray}
\label{r03-D}
i\frac{d\left<\hat{n}_{2\uparrow}\right>}{dT}=
-t_{1\uparrow}\left<\hat{n}_{12\uparrow}\right>+t_{2\uparrow}\left<\hat{n}_{21\uparrow}\right>
+J_{2\uparrow}\left<\hat{n}_{2\uparrow\downarrow}\right>-J_{2\downarrow}\left<\hat{n}_{2\downarrow\uparrow}\right>+
P_{2}\left<\hat{\Delta}^{\dagger}_{2}\right>-P^{\star}_{2}\left<\hat{\Delta}_{2}\right>,
\end{eqnarray}
\begin{eqnarray}
\label{r04-D}
i\frac{d\left<\hat{n}_{2\downarrow}\right>}{dT}=
-t_{1\downarrow}\left<\hat{n}_{12\downarrow}\right>+t_{2\downarrow}\left<\hat{n}_{21\downarrow}\right>
+J_{2\downarrow}\left<\hat{n}_{2\downarrow\uparrow}\right>-J_{2\uparrow}\left<\hat{n}_{2\uparrow\downarrow}\right>+
P_{2}\left<\hat{\Delta}^{\dagger}_{2}\right>-P^{\star}_{2}\left<\hat{\Delta}_{2}\right>,
\end{eqnarray}
\begin{eqnarray}
\label{r05-D}
i\frac{d\left<\hat{n}_{12\uparrow}\right>}{dT}=
-\varepsilon_{1\uparrow}\left<\hat{n}_{12\uparrow}\right>+\varepsilon_{2\uparrow}\left<\hat{n}_{12\uparrow}\right>
+t_{2\uparrow}\left<\hat{n}_{1\uparrow}\right>-t_{2\uparrow}\left<\hat{n}_{2\uparrow}\right>,
\end{eqnarray}
\begin{eqnarray}
\label{r06-D}
i\frac{d\left<\hat{n}_{12\downarrow}\right>}{dT}=
-\varepsilon_{1\downarrow}\left<\hat{n}_{12\downarrow}\right>+\varepsilon_{2\downarrow}\left<\hat{n}_{12\downarrow}\right>
+t_{2\downarrow}\left<\hat{n}_{1\downarrow}\right>-t_{2\downarrow}\left<\hat{n}_{2\downarrow}\right>,
\end{eqnarray}
\begin{eqnarray}
\label{r07-D}
i\frac{d\left<\hat{n}_{21\uparrow}\right>}{dT}=
\varepsilon_{1\uparrow}\left<\hat{n}_{21\uparrow}\right>-\varepsilon_{2\uparrow}\left<\hat{n}_{21\uparrow}\right>
+t_{1\uparrow}\left<\hat{n}_{2\uparrow}\right>-t_{1\uparrow}\left<\hat{n}_{1\uparrow}\right>,
\end{eqnarray}
\begin{eqnarray}
\label{r08-D}
i\frac{d\left<\hat{n}_{21\downarrow}\right>}{dT}=
\varepsilon_{1\downarrow}\left<\hat{n}_{21\downarrow}\right>-\varepsilon_{2\downarrow}\left<\hat{n}_{21\downarrow}\right>
+t_{1\downarrow}\left<\hat{n}_{2\downarrow}\right>-t_{1\downarrow}\left<\hat{n}_{1\downarrow}\right>,
\end{eqnarray}
\begin{eqnarray}
\label{r09-D}
i\frac{d\left<\hat{n}_{1\uparrow\downarrow}\right>}{dT}=
-\varepsilon_{1\uparrow}\left<\hat{n}_{1\uparrow\downarrow}\right>+\varepsilon_{1\downarrow}\left<\hat{n}_{1\uparrow\downarrow}\right>
+J_{1\downarrow}\left<\hat{n}_{1\uparrow}\right>-J_{1\downarrow}\left<\hat{n}_{1\downarrow}\right>,
\end{eqnarray}
\begin{eqnarray}
\label{r10-D}
i\frac{d\left<\hat{n}_{1\downarrow\uparrow}\right>}{dT}=
\varepsilon_{1\uparrow}\left<\hat{n}_{1\downarrow\uparrow}\right>-\varepsilon_{1\downarrow}\left<\hat{n}_{1\downarrow\uparrow}\right>
+J_{1\uparrow}\left<\hat{n}_{1\downarrow}\right>-J_{1\uparrow}\left<\hat{n}_{1\uparrow}\right>,
\end{eqnarray}
\begin{eqnarray}
\label{r11-D}
i\frac{d\left<\hat{n}_{2\uparrow\downarrow}\right>}{dT}=
-\varepsilon_{2\uparrow}\left<\hat{n}_{2\uparrow\downarrow}\right>+\varepsilon_{2\downarrow}\left<\hat{n}_{2\uparrow\downarrow}\right>
+J_{2\downarrow}\left<\hat{n}_{2\uparrow}\right>-J_{2\downarrow}\left<\hat{n}_{2\downarrow}\right>,
\end{eqnarray}
\begin{eqnarray}
\label{r12-D}
i\frac{d\left<\hat{n}_{2\downarrow\uparrow}\right>}{dT}=
\varepsilon_{2\uparrow}\left<\hat{n}_{2\downarrow\uparrow}\right>-\varepsilon_{2\downarrow}\left<\hat{n}_{2\downarrow\uparrow}\right>
+J_{2\uparrow}\left<\hat{n}_{2\downarrow}\right>-J_{2\uparrow}\left<\hat{n}_{2\uparrow}\right>,
\end{eqnarray}
\begin{eqnarray}
\label{r13-D}
i\frac{d\left<\hat{\Delta}^{\dagger}_{1}\right>}{dT}=
-\varepsilon_{1\uparrow}\left<\hat{\Delta}^{\dagger}_{1}\right>-\varepsilon_{1\downarrow}\left<\hat{\Delta}^{\dagger}_{1}\right>
+P^{\star}_{1}\left<\hat{n}_{1\uparrow}\right>+P^{\star}_{1}\left<\hat{n}_{1\downarrow}\right>-P^{\star}_{1},
\end{eqnarray}
\begin{eqnarray}
\label{r14-D}
i\frac{d\left<\hat{\Delta}_{1}\right>}{dT}=
\varepsilon_{1\uparrow}\left<\hat{\Delta}_{1}\right>+\varepsilon_{1\downarrow}\left<\hat{\Delta}_{1}\right>
-P_{1}\left<\hat{n}_{1\downarrow}\right>-P_{1}\left<\hat{n}_{1\uparrow}\right>+P_{1},
\end{eqnarray}
\begin{eqnarray}
\label{r15-D}
i\frac{d\left<\hat{\Delta}^{\dagger}_{2}\right>}{dT}=
-\varepsilon_{2\uparrow}\left<\hat{\Delta}^{\dagger}_{2}\right>-\varepsilon_{2\downarrow}\left<\hat{\Delta}^{\dagger}_{2}\right>
+P^{\star}_{2}\left<\hat{n}_{2\uparrow}\right>+P^{\star}_{2}\left<\hat{n}_{2\downarrow}\right>-P^{\star}_{2},
\end{eqnarray}
\begin{eqnarray}
\label{r16-D}
i\frac{d\left<\hat{\Delta}_{2}\right>}{dT}=
\varepsilon_{2\uparrow}\left<\hat{\Delta}_{2}\right>+\varepsilon_{2\downarrow}\left<\hat{\Delta}_{2}\right>
-P_{2}\left<\hat{n}_{2\downarrow}\right>-P_{2}\left<\hat{n}_{2\uparrow}\right>+P_{2}.
\end{eqnarray}

\newpage
\section{\label{dodE} The system of differential equations for electron observables ($\gamma\neq 0$)}

The system of differential equations can be written in the form:
\begin{eqnarray}
\label{r01-E}
i\frac{d\left<\hat{n}_{1\uparrow}\right>}{dT}&=&
t_{1\uparrow}\left<\hat{n}_{12\uparrow}\right>-t_{2\uparrow}\left<\hat{n}_{21\uparrow}\right>
+J_{1\uparrow}\left<\hat{n}_{1\uparrow\downarrow}\right>-J_{1\downarrow}\left<\hat{n}_{1\downarrow\uparrow}\right>+
P_{1}\left<\hat{\Delta}^{\dagger}_{1}\right>-P^{\star}_{1}\left<\hat{\Delta}_{1}\right>\\ \nonumber
&+&i\gamma\left<\hat{n}_{1\uparrow}\right>-2i\gamma|\Delta_{1}|^{2}\\ \nonumber
&-&2i\gamma\left<\hat{n}_{1\uparrow}\right>\sum_{\sigma}\left(\left<\hat{n}_{1\sigma}\right>-\left<\hat{n}_{2\sigma}\right>\right),
\end{eqnarray}
\begin{eqnarray}
\label{r02-E}
i\frac{d\left<\hat{n}_{1\downarrow}\right>}{dT}&=&
t_{1\downarrow}\left<\hat{n}_{12\downarrow}\right>-t_{2\downarrow}\left<\hat{n}_{21\downarrow}\right>
+J_{1\downarrow}\left<\hat{n}_{1\downarrow\uparrow}\right>-J_{1\uparrow}\left<\hat{n}_{1\uparrow\downarrow}\right>+
P_{1}\left<\hat{\Delta}^{\dagger}_{1}\right>-P^{\star}_{1}\left<\hat{\Delta}_{1}\right>\\ \nonumber
&+&i\gamma\left<\hat{n}_{1\downarrow}\right>-2i\gamma|\Delta_{1}|^{2}\\ \nonumber
&-&2i\gamma\left<\hat{n}_{1\downarrow}\right>\sum_{\sigma}\left(\left<\hat{n}_{1\sigma}\right>-\left<\hat{n}_{2\sigma}\right>\right),
\end{eqnarray}
\begin{eqnarray}
\label{r03-E}
i\frac{d\left<\hat{n}_{2\uparrow}\right>}{dT}&=&
-t_{1\uparrow}\left<\hat{n}_{12\uparrow}\right>+t_{2\uparrow}\left<\hat{n}_{21\uparrow}\right>
+J_{2\uparrow}\left<\hat{n}_{2\uparrow\downarrow}\right>-J_{2\downarrow}\left<\hat{n}_{2\downarrow\uparrow}\right>+
P_{2}\left<\hat{\Delta}^{\dagger}_{2}\right>-P^{\star}_{2}\left<\hat{\Delta}_{2}\right>\\ \nonumber
&-&i\gamma\left<\hat{n}_{2\uparrow}\right>+2i\gamma|\Delta_{2}|^{2}\\ \nonumber
&-&2i\gamma\left<\hat{n}_{2\uparrow}\right>\sum_{\sigma}\left(\left<\hat{n}_{1\sigma}\right>-\left<\hat{n}_{2\sigma}\right>\right),
\end{eqnarray}
\begin{eqnarray}
\label{r04-E}
i\frac{d\left<\hat{n}_{2\downarrow}\right>}{dT}&=&
-t_{1\downarrow}\left<\hat{n}_{12\downarrow}\right>+t_{2\downarrow}\left<\hat{n}_{21\downarrow}\right>
+J_{2\downarrow}\left<\hat{n}_{2\downarrow\uparrow}\right>-J_{2\uparrow}\left<\hat{n}_{2\uparrow\downarrow}\right>+
P_{2}\left<\hat{\Delta}^{\dagger}_{2}\right>-P^{\star}_{2}\left<\hat{\Delta}_{2}\right>\\ \nonumber
&-&i\gamma\left<\hat{n}_{2\downarrow}\right>+2i\gamma|\Delta_{2}|^{2}\\ \nonumber
&-&2i\gamma\left<\hat{n}_{2\downarrow}\right>\sum_{\sigma}\left(\left<\hat{n}_{1\sigma}\right>-\left<\hat{n}_{2\sigma}\right>\right),
\end{eqnarray}
\begin{eqnarray}
\label{r05-E}
i\frac{d\left<\hat{n}_{12\uparrow}\right>}{dT}&=&
-\varepsilon_{1\uparrow}\left<\hat{n}_{12\uparrow}\right>+\varepsilon_{2\uparrow}\left<\hat{n}_{12\uparrow}\right>
+t_{2\uparrow}\left<\hat{n}_{1\uparrow}\right>-t_{2\uparrow}\left<\hat{n}_{2\uparrow}\right>\\ \nonumber
&+&i\gamma\left<\hat{n}_{12\uparrow}\right>\\ \nonumber
&-&2i\gamma\left<\hat{n}_{12\uparrow}\right>\sum_{\sigma}\left(\left<\hat{n}_{1\sigma}\right>-\left<\hat{n}_{2\sigma}\right>\right),
\end{eqnarray}
\begin{eqnarray}
\label{r06-E}
i\frac{d\left<\hat{n}_{12\downarrow}\right>}{dT}&=&
-\varepsilon_{1\downarrow}\left<\hat{n}_{12\downarrow}\right>+\varepsilon_{2\downarrow}\left<\hat{n}_{12\downarrow}\right>
+t_{2\downarrow}\left<\hat{n}_{1\downarrow}\right>-t_{2\downarrow}\left<\hat{n}_{2\downarrow}\right>\\ \nonumber
&+&i\gamma\left<\hat{n}_{12\downarrow}\right>\\ \nonumber
&-&2i\gamma\left<\hat{n}_{12\downarrow}\right>\sum_{\sigma}\left(\left<\hat{n}_{1\sigma}\right>-\left<\hat{n}_{2\sigma}\right>\right),
\end{eqnarray}
\begin{eqnarray}
\label{r07-E}
i\frac{d\left<\hat{n}_{21\uparrow}\right>}{dT}&=&
\varepsilon_{1\uparrow}\left<\hat{n}_{21\uparrow}\right>-\varepsilon_{2\uparrow}\left<\hat{n}_{21\uparrow}\right>
+t_{1\uparrow}\left<\hat{n}_{2\uparrow}\right>-t_{1\uparrow}\left<\hat{n}_{1\uparrow}\right>\\ \nonumber
&-&i\gamma\left<\hat{n}_{21\uparrow}\right>\\ \nonumber
&-&2i\gamma\left<\hat{n}_{21\uparrow}\right>\sum_{\sigma}\left(\left<\hat{n}_{1\sigma}\right>-\left<\hat{n}_{2\sigma}\right>\right),
\end{eqnarray}
\begin{eqnarray}
\label{r08-E}
i\frac{d\left<\hat{n}_{21\downarrow}\right>}{dT}&=&
\varepsilon_{1\downarrow}\left<\hat{n}_{21\downarrow}\right>-\varepsilon_{2\downarrow}\left<\hat{n}_{21\downarrow}\right>
+t_{1\downarrow}\left<\hat{n}_{2\downarrow}\right>-t_{1\downarrow}\left<\hat{n}_{1\downarrow}\right>\\ \nonumber
&-&i\gamma\left<\hat{n}_{21\downarrow}\right>\\ \nonumber
&-&2i\gamma\left<\hat{n}_{21\downarrow}\right>\sum_{\sigma}\left(\left<\hat{n}_{1\sigma}\right>-\left<\hat{n}_{2\sigma}\right>\right),
\end{eqnarray}
\begin{eqnarray}
\label{r09-E}
i\frac{d\left<\hat{n}_{1\uparrow\downarrow}\right>}{dT}&=&
-\varepsilon_{1\uparrow}\left<\hat{n}_{1\uparrow\downarrow}\right>+\varepsilon_{1\downarrow}\left<\hat{n}_{1\uparrow\downarrow}\right>
+J_{1\downarrow}\left<\hat{n}_{1\uparrow}\right>-J_{1\downarrow}\left<\hat{n}_{1\downarrow}\right>\\ \nonumber
&+&i\gamma\left<\hat{n}_{1\uparrow\downarrow}\right>\\ \nonumber
&-&2i\gamma\left<\hat{n}_{1\uparrow\downarrow}\right>\sum_{\sigma}\left(\left<\hat{n}_{1\sigma}\right>-\left<\hat{n}_{2\sigma}\right>\right),
\end{eqnarray}
\begin{eqnarray}
\label{r10-E}
i\frac{d\left<\hat{n}_{1\downarrow\uparrow}\right>}{dT}&=&
\varepsilon_{1\uparrow}\left<\hat{n}_{1\downarrow\uparrow}\right>-\varepsilon_{1\downarrow}\left<\hat{n}_{1\downarrow\uparrow}\right>
+J_{1\uparrow}\left<\hat{n}_{1\downarrow}\right>-J_{1\uparrow}\left<\hat{n}_{1\uparrow}\right>\\ \nonumber
&+&i\gamma\left<\hat{n}_{1\downarrow\uparrow}\right>\\ \nonumber
&-&2i\gamma\left<\hat{n}_{1\downarrow\uparrow}\right>\sum_{\sigma}\left(\left<\hat{n}_{1\sigma}\right>-\left<\hat{n}_{2\sigma}\right>\right),
\end{eqnarray}
\begin{eqnarray}
\label{r11-E}
i\frac{d\left<\hat{n}_{2\uparrow\downarrow}\right>}{dT}&=&
-\varepsilon_{2\uparrow}\left<\hat{n}_{2\uparrow\downarrow}\right>+\varepsilon_{2\downarrow}\left<\hat{n}_{2\uparrow\downarrow}\right>
+J_{2\downarrow}\left<\hat{n}_{2\uparrow}\right>-J_{2\downarrow}\left<\hat{n}_{2\downarrow}\right>\\ \nonumber
&-&i\gamma\left<\hat{n}_{2\uparrow\downarrow}\right>\\ \nonumber
&-&2i\gamma\left<\hat{n}_{2\uparrow\downarrow}\right>\sum_{\sigma}\left(\left<\hat{n}_{1\sigma}\right>-\left<\hat{n}_{2\sigma}\right>\right),
\end{eqnarray}
\begin{eqnarray}
\label{r12-E}
i\frac{d\left<\hat{n}_{2\downarrow\uparrow}\right>}{dT}&=&
\varepsilon_{2\uparrow}\left<\hat{n}_{2\downarrow\uparrow}\right>-\varepsilon_{2\downarrow}\left<\hat{n}_{2\downarrow\uparrow}\right>
+J_{2\uparrow}\left<\hat{n}_{2\downarrow}\right>-J_{2\uparrow}\left<\hat{n}_{2\uparrow}\right>\\ \nonumber
&-&i\gamma\left<\hat{n}_{2\downarrow\uparrow}\right>\\ \nonumber
&-&2i\gamma\left<\hat{n}_{2\downarrow\uparrow}\right>\sum_{\sigma}\left(\left<\hat{n}_{1\sigma}\right>-\left<\hat{n}_{2\sigma}\right>\right),
\end{eqnarray}
\begin{eqnarray}
\label{r13-E}
i\frac{d\left<\hat{\Delta}^{\dagger}_{1}\right>}{dT}&=&
-\varepsilon_{1\uparrow}\left<\hat{\Delta}^{\dagger}_{1}\right>-\varepsilon_{1\downarrow}\left<\hat{\Delta}^{\dagger}_{1}\right>
+P^{\star}_{1}\left<\hat{n}_{1\uparrow}\right>+P^{\star}_{1}\left<\hat{n}_{1\downarrow}\right>-P^{\star}_{1}\\ \nonumber
&+&2i\gamma\left<\hat{\Delta}^{\dagger}_{1}\right>+2i\gamma\left<\hat{\Delta}^{\dagger}_{1}\right>\left<\hat{n}_{1\downarrow}\right>\\ \nonumber
&-&2i\gamma\left<\hat{\Delta}^{\dagger}_{1}\right>\sum_{\sigma}\left(\left<\hat{n}_{1\sigma}\right>-\left<\hat{n}_{2\sigma}\right>\right),
\end{eqnarray}
\begin{eqnarray}
\label{r14-E}
i\frac{d\left<\hat{\Delta}_{1}\right>}{dT}&=&
\varepsilon_{1\uparrow}\left<\hat{\Delta}_{1}\right>+\varepsilon_{1\downarrow}\left<\hat{\Delta}_{1}\right>
-P_{1}\left<\hat{n}_{1\downarrow}\right>-P_{1}\left<\hat{n}_{1\uparrow}\right>+P_{1}\\ \nonumber
&-&i\gamma\left<\hat{\Delta}_{1}\right>+2i\gamma\left<\hat{\Delta}_{1}\right>\left<\hat{n}_{1\downarrow}\right>\\ \nonumber
&-&2i\gamma\left<\hat{\Delta}_{1}\right>\sum_{\sigma}\left(\left<\hat{n}_{1\sigma}\right>-\left<\hat{n}_{2\sigma}\right>\right),
\end{eqnarray}
\begin{eqnarray}
\label{r15-E}
i\frac{d\left<\hat{\Delta}^{\dagger}_{2}\right>}{dT}&=&
-\varepsilon_{2\uparrow}\left<\hat{\Delta}^{\dagger}_{2}\right>-\varepsilon_{2\downarrow}\left<\hat{\Delta}^{\dagger}_{2}\right>
+P^{\star}_{2}\left<\hat{n}_{2\uparrow}\right>+P^{\star}_{2}\left<\hat{n}_{2\downarrow}\right>-P^{\star}_{2}\\ \nonumber
&-&2i\gamma\left<\hat{\Delta}^{\dagger}_{2}\right>-2i\gamma\left<\hat{\Delta}^{\dagger}_{2}\right>\left<\hat{n}_{2\downarrow}\right>\\ \nonumber
&-&2i\gamma\left<\hat{\Delta}^{\dagger}_{2}\right>\sum_{\sigma}\left(\left<\hat{n}_{1\sigma}\right>-\left<\hat{n}_{2\sigma}\right>\right),
\end{eqnarray}
\begin{eqnarray}
\label{r16-E}
i\frac{d\left<\hat{\Delta}_{2}\right>}{dT}&=&
\varepsilon_{2\uparrow}\left<\hat{\Delta}_{2}\right>+\varepsilon_{2\downarrow}\left<\hat{\Delta}_{2}\right>
-P_{2}\left<\hat{n}_{2\downarrow}\right>-P_{2}\left<\hat{n}_{2\uparrow}\right>+P_{2}\\ \nonumber
&+&i\gamma\left<\hat{\Delta}_{2}\right>-2i\gamma\left<\hat{\Delta}_{2}\right>\left<\hat{n}_{2\downarrow}\right>\\ \nonumber
&-&2i\gamma\left<\hat{\Delta}_{2}\right>\sum_{\sigma}\left(\left<\hat{n}_{1\sigma}\right>-\left<\hat{n}_{2\sigma}\right>\right).
\end{eqnarray}
%

\bibliography{Bibliography}
\end{document}